\documentclass[%
 aip,
 amsmath,amssymb,
reprint, onecolumn 
]{revtex4-1}

\usepackage[]{standalone}
\usepackage{import}
\usepackage[utf8]{inputenc}
\usepackage[T1]{fontenc}
\usepackage{cmap}
\usepackage{graphicx}
\usepackage{mathtools}

\usepackage[hyperfootnotes=false,breaklinks=true]{hyperref}
\usepackage{setspace}
\hypersetup{
 bookmarksopen=true,
 bookmarksopenlevel=1,
 colorlinks=true,
 linkcolor=blue,
 anchorcolor=blue,
 citecolor=blue,
 filecolor=blue,
 urlcolor=blue,
 pdfpagemode=UseOutlines,
 pdfstartview={XYZ null null 1},
 linktocpage=true,
}
\usepackage{amsmath}
\allowdisplaybreaks  
\usepackage{amssymb}
\usepackage{amstext}
\usepackage{amsfonts}
\usepackage{mathrsfs}
\usepackage{amsthm}
\usepackage{dsfont}
\usepackage{bm}
\usepackage{braket}
\usepackage{physics}
\usepackage{enumitem}
\usepackage[
 capitalise,
]{cleveref}

\usepackage{pgfplots}
\pgfplotsset{compat=newest}
\usepackage{tikz}
\usetikzlibrary{
  positioning,
  shapes.geometric,
  arrows,
  arrows.meta,
  backgrounds,
  fit,
  intersections,  
}
\usepackage[english]{babel}
\usepackage{qcircuit} 
\usepackage{algorithm,algorithmic}

\usepackage{comment}

\newcommand{\subfig}[2]{%
    {#1} \vtop{
  \vskip0pt
  \hbox{#2}
}}

\theoremstyle{plain}

\theoremstyle{definition}

\newcommand{\mycomment}[1]{}

\usepackage{relsize}
\usepackage{cancel}
\usepackage{amsthm}
\newtheoremstyle{nopunct}
  {3pt}                
  {3pt}                
  {\upshape}           
  {}                   
  {\bfseries}          
  {:}                   
  { }                  
  {}                   
\newtheorem{definition}{Definition}
\theoremstyle{nopunct}

\makeatletter
\def\blfootnote{\xdef\@thefnmark{}\@footnotetext}

\makeatletter
\def\@email#1#2{%
 \endgroup
 \patchcmd{\titleblock@produce}
  {\frontmatter@RRAPformat}
  {\frontmatter@RRAPformat{\produce@RRAP{*#1\href{mailto:#2}{#2}}}\frontmatter@RRAPformat}
  {}{}
}%
\makeatother
\begin{document}


\title{Locally Purified Maximally Mixed States At Scale: \\ Entanglement Pruning and Symmetries}

\author{Amit Jamadagni Gangapuram}  
\email{gangapurama@ornl.gov}
\affiliation{Computational Sciences and Engineering Division, %
Oak Ridge National Laboratory, %
Oak Ridge, Tennessee 37831, USA}

\author{Eugene Dumitrescu}  
\email{dumitrescuef@ornl.gov}
\affiliation{Computational Sciences and Engineering Division, %
Oak Ridge National Laboratory, %
Oak Ridge, Tennessee 37831, USA}

\begin{abstract}
Locally Purified Density Operators (LPDOs) are state-of-the-art tensor network ansatze candidates that efficiently represent mixed quantum states at scale. However, given their non-uniqueness, their representational complexity is generally sub-optimal in practical computations. In this work we perform a comprehensive numerical and analytical analysis and resolve this issue in the experimentally relevant limit where noise depolarizes the density operator into a maximally mixed state. To resolve the sub-optimality issue, we analyze two numerical tools, one analytic method, and detail the relations between them. The numerical tools used are fidelity-preserving truncations and isometric gauge transformations leveraging Riemannian optimizations over entropic objective functions. In addition, by invoking the injectivity and symmetry constraints of the maximally mixed LPDO, we also present analytical closed-form expressions for the disentangler and discuss their relation to numerical optimizers. Further, away from the maximally mixed state, our simulations highlight how the truncation threshold smoothly interpolate, as a function of depolarization, between established matrix product results and our new results. Our work shows how, by minimizing the resources required to represent key states of practical interest in experiment, the efficiency of tensor network algorithms can be substantially increased. This paves the path for uncovering tensor network's fundamental scalability limits and latent potential in representing the wide locus of mixed quantum states that are accessible on near-term quantum devices. 
\end{abstract}
\maketitle



\section{Introduction}

The identity operator is a central object in arithmetic, group, and category theories in mathematics. It plays an equally important role in physical theories such as quantum mechanics, where it is the Casimir operator and where the resolution of identity often aids both analytic proofs and numerical techniques. In addition, the normalized identity operator represents infinite temperature states as density operators. 

The understanding of these states remains key in condensed matter physics as they represent one limit of phase diagrams (with respect to temperature). In fact, there has been a growing interest in studying mixed state phase transitions~\cite{Chen2024,Ellison2025,Guo2025} as well as representing finite temperature states, in both the presence and absence of symmetries. In this context, interesting counter intuitive phenomena such as maximally mixed but entangled states within symmetry sectors have been reported~\cite{Moharramipour2024}.  On a more practical front, this is the state that polynomial depth quantum computing algorithms converge to on pre-fault-tolerant quantum computing hardware. Indeed, it is so easy to produce these states in quantum computations that the rate of convergence to this state, and extrapolation back to the noiseless result, has emerged as the most powerful and scalable error mitigation technique in today's quantum computations~\cite{Temme2017, Li2017, Kandala2018, Cai2021,Frey2022}. Recent experiments~\cite{Frey2022,Arute2019} claiming quantum advantage have reported mixed states that are close to the maximally mixed state. Hence understanding representation of such states remains vital for gaining insight into quantum computational advantage~\cite{Zhou2020}.

From a representation theory viewpoint, tensor network methods, such as the matrix product states (MPS)~\cite{Perez2007} and generalizations~\cite{Vidal2007,Verstraete2004}, are extremely useful in efficiently providing $\epsilon$-approximations of entangled pure quantum states. These results motivate the question of whether, in analogy with pure states, similar approximations of mixed states are possible~\cite{Cuevas2013,Jaschke2019,Cuevas2020}. Locally Purified Density Operators (LPDOs) are one leading candidate~\cite{Verstraete2004_mpdo,Guo2022,Mueller2024,Wanisch2025,Harada2025,Guo2024} to efficiently represent \textit{mixed} quantum states. Along with matrix product operators (MPOs), LPDOs are a natural extension of the MPS canonical form. This is because they encode both coherent as well as classical mixture correlations. Further unlike the MPO, the LPDO manifestly enforces the density operator's positivity constraint (see App.~\ref{app:review_lpdo} for details), while also providing an explicit handle on classical correlations and a quadratic reduction in the dimension of the virtual entanglement-mediating index. While the tensor mechanics of noise channels and unitaries on LPDOs has been extensively discussed~\cite{Werner2016,Jamadagni2024}, for a more exhaustive list, we refer the reader to the table of equivalent representations in Ref.~\citenum{Mueller2024}, their efficiency in experimentally relevant limits remains largely unexplored. 

\begin{figure}
    \centering
    \includegraphics[width=0.6\linewidth]{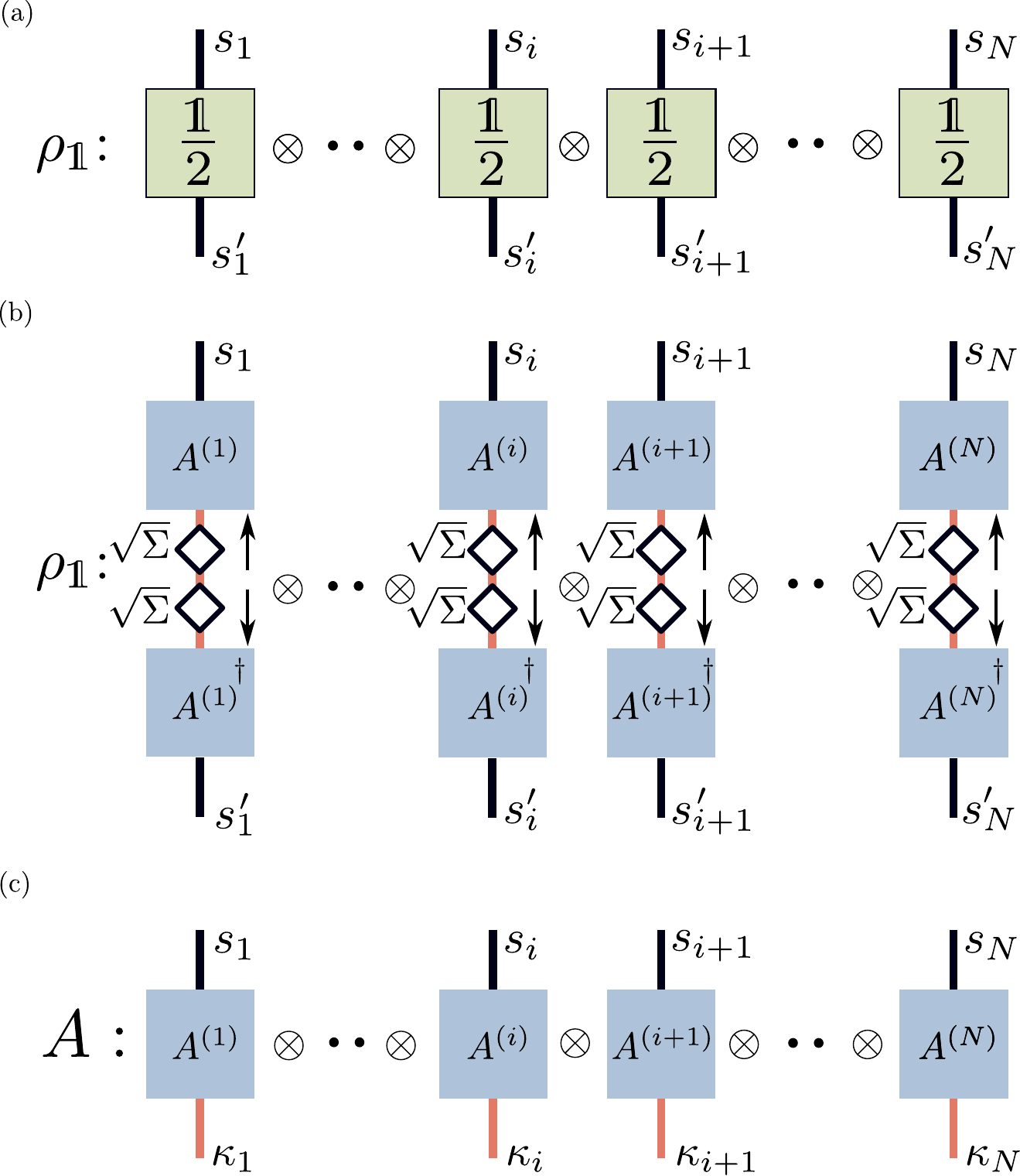}
    \caption{Optimal representation of LPMM$\rho$. (a) MPO representation of $\rho_{\mathds{1}}$. The system is completely separable, therefore $\chi=1$ between each of the neighoring tensors. (b) Performing an SVD operation on the individual tensors results in the singular values $\Sigma$, as in Eq.~\ref{eq:A_j}. The square root of
   $\Sigma$ is then merged into $A^{(i)}$ and ${A^{(i)}}^{\dag}$ at all sites resulting in the canonical LPDO.  (c) This generates the purification index $\kappa$ that encodes the mixture correlations. In the context of the 
    optimal LPMM$\rho$, we note that $\chi_{i}=1$ and $\kappa_{i}=2$ at all sites.}
    \label{fig:noise_cartoon_opt}
\end{figure}

In a previous work~\cite{Jamadagni2024} the authors discovered that, despite its conceptual simplicity, the LPDO representation of the maximally mixed state was \textit{dramatically} sub-optimal. Optimizing the representation of the completely depolarized state as it arises in practical computations is therefore crucial. One must first understand when efficient representations are possible and subsequently how much work is required to constructively find them. Since partially depolarized states $\rho = (1-\epsilon) \rho_\mathds{1} + \epsilon \sigma$, where $\sigma$ is a non-separable density operator, are natural generalizations systematically diverging from this complex but trivial fixed point, the answer to these questions has tremendous implications for pushing the fundamental limits of classical approaches to quantum simulations and qubit modeling. 

Therefore, in this work, we revisit this issue and address it by introducing a variety of numerical and analytic tools. We show how these tools are related and discuss applications of these tools for the representation, with a minimal resource cost, of generic density operators which are sampled in near-term quantum computing architectures. In Sec.~\ref{sec:den_op_rep}, we begin by presenting equivalent LPDO representations, an optimal and sub-optimal version, of the maximally mixed state. Next, in Sec.~\ref{sec:rep_tools}, we introduce different numerical and analytical tools that map the sub-optimal representation to its optimal counterpart. Further, in Sec.~\ref{sec:away_from_mms}, we quantify the effectiveness of the introduced numerical techniques on partially depolarized density operators. Finally in Sec.~\ref{sec:conclusion}, we summarize the results while also providing few directions that can be further explored using the methods introduced.

\begin{figure}
    \centering
    \includegraphics[width=0.6\linewidth]{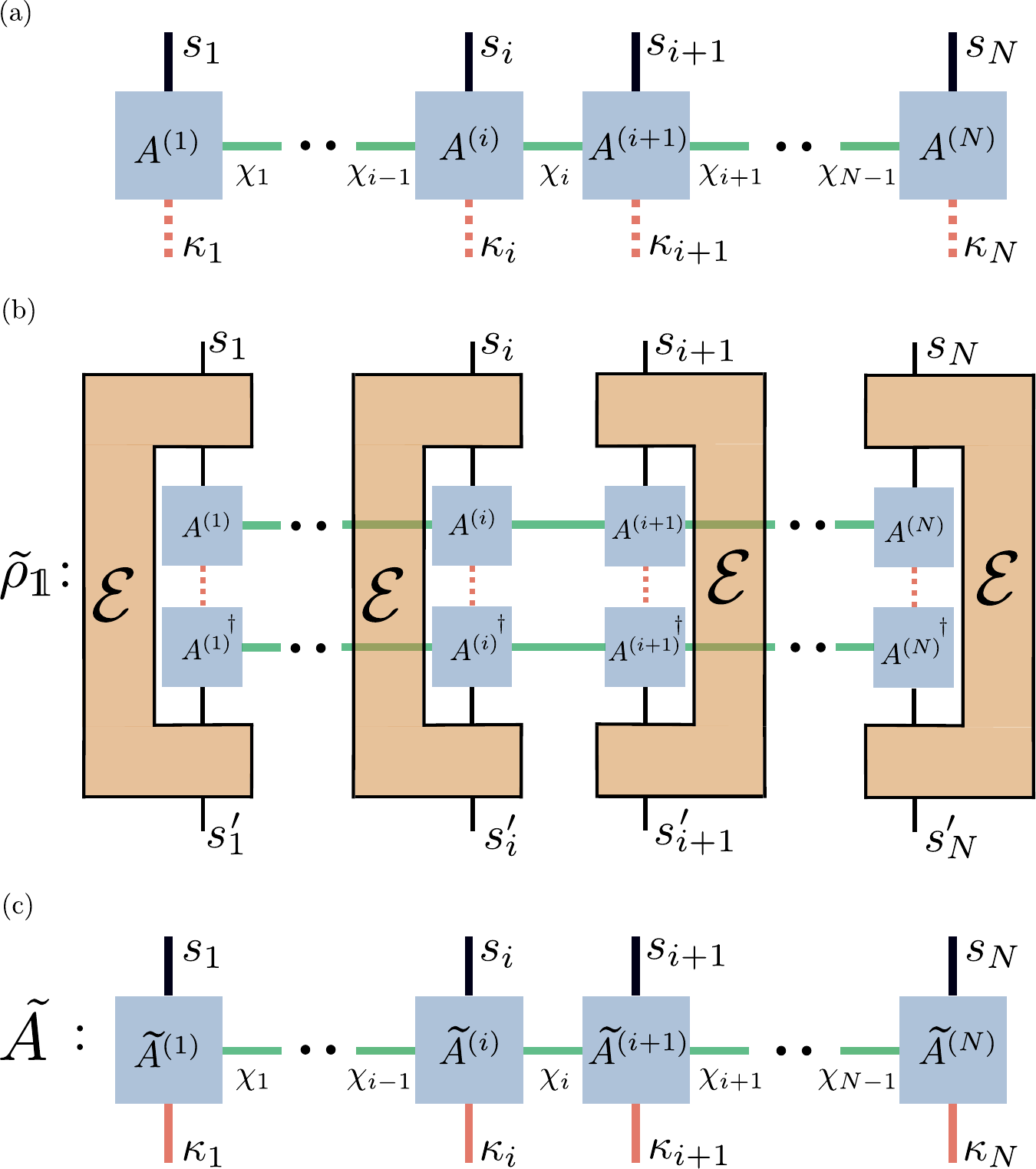}
    \caption{Sub-optimal representation of LPMM$\rho$. (a) Initializing a pure random MPS of $N$ sites with 
    $\chi_{i}$'s bounded by $\chi_{\text{max}}$ by employing the \text{randomMPS} function call in ITensors.jl. 
    We decorate the above with the mixture indices, $\kappa_{i}=1$ (denoted by dashed orange line) at each $i$ as the 
    state remains pure. (b) The action of bitflip and dephasing single qubit noise channels at each of the sites, for a 
    more detailed description of the above, see App.~\ref{appa:channel_application}. The 
    noise channels completely decohere the initial entangled state, resulting in a LPMM$\rho$. 
    (c) Representationally, $\kappa_{i}$'s are inflated due to the noise channels (denoted by solid orange line), 
    however $\chi$ remains unchanged leading to a sub-optimal representation of LPMM$\rho$.}
    \label{fig:noise_cartoon_sub_opt}
\end{figure}

\section{Density Operators and Representation by Local Purification} \label{sec:den_op_rep}

We begin by introducing the maximally mixed state and the corresponding normalized density operator. Consider a system of $N$-qudits, i.e., quantum systems with $D$-dimensional basis states $\mathcal{B} = \{\ket{0}, \ket{1}, \cdots, \ket{D-1}\}$. The corresponding identity and normalized density operator are expressed as
\begin{align}
  \mathds{1} &= \bigotimes_{n=1}^{N} \sum_{q=0}^{D-1} \ket{q}_{n}\bra{q}  \\
  \rho_\mathds{1} &= \bigotimes_{n=1}^{N} \frac{1}{D} \sum_{q=0}^{D-1}\ket{q}_{n}\bra{q}.
\end{align}
Alternatively, in a possibly entangled $D$-dimensional energy eigenbasis--with eigen energy eigenvector pairs denoted as $(\lambda, \ket{\lambda})$, we have 
\begin{align}
\mathds{1} &= \sum_{\lambda} \ketbra{\lambda}{\lambda} \\
\rho_\mathds{1} &= \frac{1}{D} \sum_{\lambda} \ketbra{\lambda}{\lambda}.    
\label{eq:rho_MM}
\end{align}
When represented via the locally purified tensor network ansatze, we refer to the above maximally mixed state as the Locally Purified Maximally Mixed Density Operator (LPMM$\rho$). In the next section, we present two different strategies that prepare two distinct representations of the LPMM$\rho$.

\subsection{Background and Formalism}
To begin our study, we will first describe the optimal, i.e. minimal complexity representation of the LPMM$\rho$, which minimizes memory allocation resources. Afterwards, we will review a protocol that generates sub-optimal representation of the LPMM$\rho$ i.e., a representation that demands additional memory resources in comparison to the optimal representation. The main goal of the article is to then overcome this challenge of mapping the sub-optimal representation to the optimal representation. 
Inspired by machine learning approaches to minimizing the graph and parameter complexity of neural networks, we refer to this process as (representation-theoretic) entanglement pruning~\cite{Surace2019,Kevin2021,Cichy2025}. 

In the context of the $N$-qubit system (without loss of generality, we consider quantum states comprised of qubits; however, it is straightforward to extend the analysis to qudit systems), the infinite temperature state is given by $\rho_{\mathds{1}} = \otimes_{j=1}^{N}\frac{\mathds{1}_{j}}{2}$ where $\mathds{1}_{j} = \ket{0}_{j}\bra{0} + \ket{1}_{j}\bra{1}$ is the identity operator at site $j$. This state can be directly interpreted as an MPO and as an LPDO. Since it is a tensor product state, $\rho_{\mathds{1}}$ is explicitly a matrix product operator with $\chi_i=1$. That is, as illustrated in Fig.~\ref{fig:noise_cartoon_opt}(a) this separable and unentangled state has a trivial virtual bond dimension connecting the site tensors $i$ and $i+1$, since there are no entanglement correlations between their physical degrees of freedom. To map this to an LPDO, we perform SVD at each site, $j$ resulting in 
\begin{equation}\label{eq:A_j}
\begin{split}
\frac{\mathds{1}_{j}}{2} &= U \Sigma V^T  = \mathds{1} (\frac{1}{2} \mathds{1}) \mathds{1} \\
& = \frac{\mathds{1}}{\sqrt{2}} \frac{\mathds{1}}{\sqrt{2}} = A^{(j)}{A^{(j)}}^{\dag}
\end{split}
\end{equation}
as in Fig.~\ref{fig:noise_cartoon_opt}(b). In the last step we have symmetrically absorbed the singlular values in to $A^{(j)}$ and $A^{(j)\dagger}$. Thus, we have $\kappa=2$ for the virtual dimension encoding mixture correlations at each site, see Fig.~\ref{fig:noise_cartoon_opt}(c). 
Since $\chi=1$ (even without additional symmetries) we call this LPDO$\rho$ representation of an $N$-qubit $\rho_{\mathds{1}}$ optimal since it scales linearly, i.e., requiring $2N$ memory (all the non zero coefficients are equal to $1/\sqrt{2}$). This is already an exponential improvement over even just keeping the diagonal in a density operator matrix (or state vector) representation where the length of the diagonal is $2^N$. 

In contrast to this optimal LPMM$\rho$ form, the LPMM$\rho$ that arises from a quantum circuit which includes noise, such that the state is eventually completely depolarized up to machine precision, is strikingly different as we now explain. To consider average case scenarios, which mimic quantum coherence correlations that arise in an ideal quantum computation, we first generate a random but pure LPDO. That is, we begin by generating a random MPS whose virtual dimension is bounded by $\chi_{\text{max}}$. Since the density operator of a pure state is $\rho = \ket{\Psi} \bra{\Psi}$, this corresponds to an LPDO with $\kappa=1$ for all site tensors, see Fig.~\ref{fig:noise_cartoon_sub_opt}(a). 
This pure and entangled LPDO, is then taken as the input state for a series of quantum channels updating the state as $\rho' = \mathcal{E}(\rho) = \sum_i K_i \rho K^\dagger_i$. 

Formally, we will describe the noisy quantum maps in terms of the Kraus operator sum representation such that $\mathcal{E}(\rho) = \sum_i K_i \rho K^\dagger_i$ where $\sum_i  K^\dagger_i K_i  = \mathds{1}$. Specifically, we will apply both dephasing (d) and bitflip (b) noise channels which are described by the Kraus operators $K_{0}^{d} = \sqrt{\gamma_{d}}Z$, $K_{1}^{d} = \sqrt{1-\gamma_{d}}\mathds{1}$ and $K_{0}^{b} = \sqrt{\gamma_{b}}X$, $K_{1}^{b} = \sqrt{1-\gamma_{b}}\mathds{1}$ respectively, where $\gamma_{d}$ and $\gamma_{b}$ are the dephasing and bitflip noise rates, with $Z$ and $X$ being the Pauli-Z and Pauli-X given by
\begin{equation}
X = 
\begin{pmatrix}
0 & 1\\
1 & 0
\end{pmatrix},\quad
Z = 
\begin{pmatrix}
1 & 0\\
0 & -1
\end{pmatrix}.
\end{equation}
The noise is maximized in the case of error rates $\gamma_{d} = \gamma_{b} = 0.5$. The application of sufficient noise results in an LPDO that is maximally mixed i.e., an LPMM$\rho$. The flow of entangled pure states towards this separable fixed point was previously studied by the authors at scale\cite{Jamadagni2024}. 

However, in this scenario, the representation is sub-optimal as $\chi$ in the updated LPMM$\rho$ is unchanged from that of the initial pure LPDO. In other words, the noise channels increase the $\kappa$ dimension at each site, this is natural and reflects the states depolarization into a classical statistical mixture. However $\chi$ remains the same as in the initial LPDO, while one might expect $\chi=1$ for states which have been disentangled by noise, see Figs.~\ref{fig:noise_cartoon_sub_opt}(b), (c). Since the fidelity of the resulting state with the known form of $\rho_\mathds{1}$ is unity, up to machine precision, we know that we have the maximally mixed state but that its representation is an inefficient one. 
In the next section, to overcome this representation-theoretic challenge, we present various techniques that prune apparent coherent correlations, thus returning an optimized representation.

\section{Tools to Gauge-Fix the LPDO\label{sec:rep_tools}}
This section presents three mutually compatible routines that transform the sub-optimal LPMM$\rho$ to its optimal counterpart. We begin with a fidelity-preserving truncation which generalizes the usual notion of truncation in the virtual coherence correlation space. Afterwards, we show how this can be augmented by optimization routines. The idea here is to, via LPDO isometries acting in the mixture subspace, explore the space of gauge choices which minimize spurious coherent correlations. Lastly, these numerical tools are supplemented by analytic results showing symmetries lead to projective representations which result in injective LPDOs. We discuss how injectivity provides a stronger (more unique) result compared to numerical optimizations.

\subsection{Fidelity-Preserving Truncation\label{sec:fpt}}
The truncated singular value decomposition (SVD), in the MPS context, is the main technical tool that is leveraged to efficiently represent entanglement area-law obeying (i.e. low-rank in terms of entanglement) quantum states. To optimize the tensors with respect to a two-body interaction, or after the application of a two-body entangling gate, two site tensors are contracted and an SVD is subsequently performed. Finally, discarding the singular values smaller than a chosen cutoff returns the optimal reduced-rank approximation with respect to the correlations between the bi-partition. Since quantum states must be normalized, this is followed by renormalization. Specifically, the $L_2$ norm of the singular values is renormalized to unity such that the square of the left/right amplitudes result in a normalized pure state. 

To explain our first set of results, we first briefly review truncation and renormalization in the context of LPDO. This two-step process occurs in both the  mixture correlations space, referred to as $\kappa$-subspace, generalizing the notion of MPS renormalization as discussed in Ref.~\onlinecite{Werner2016}, and also the space of coherent correlations, referred to as $\chi$-subspace as in the case of the MPS.

\begin{figure}
\begin{center}
\includegraphics[width=0.5\linewidth]{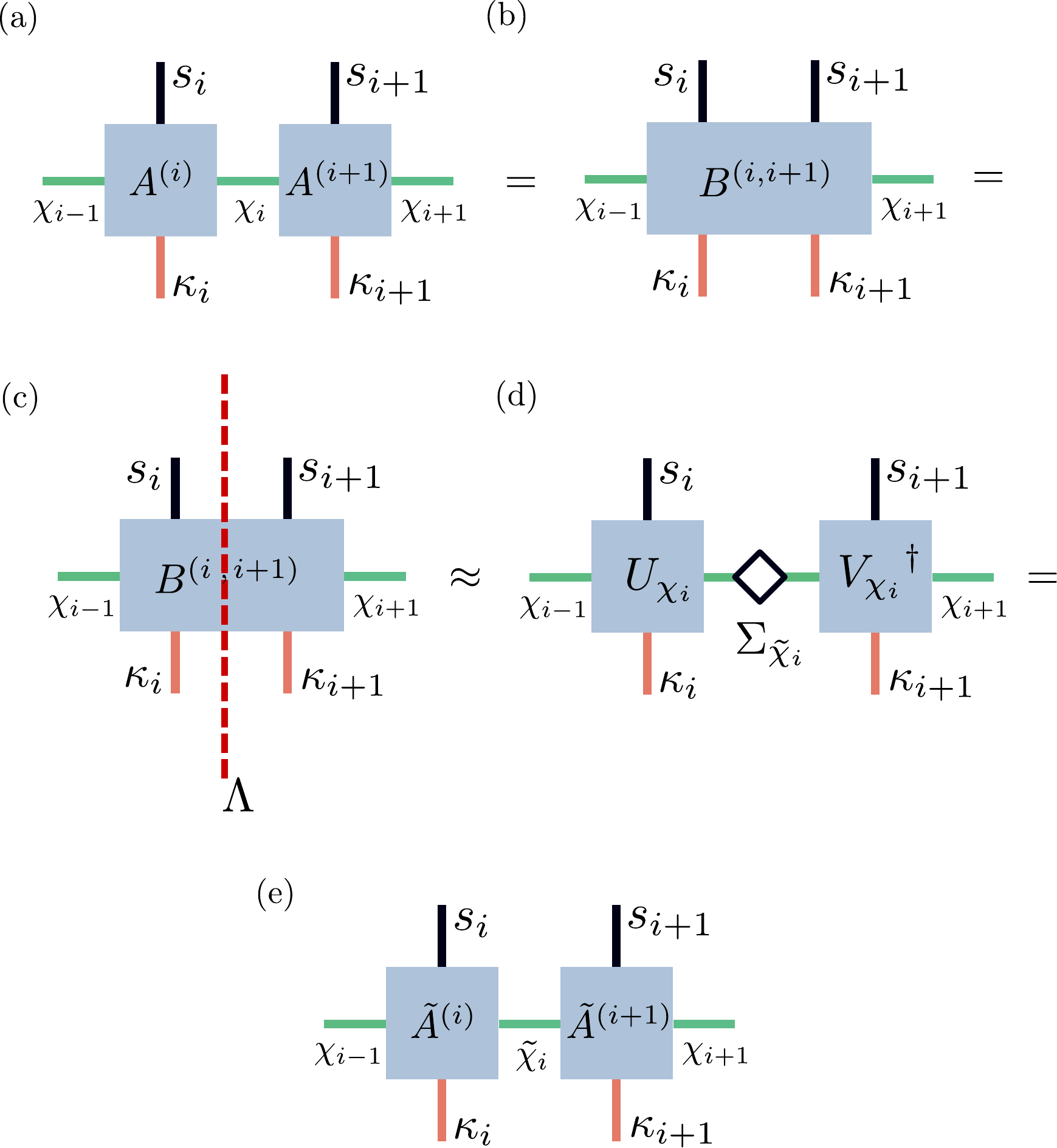}
\end{center}
\caption{Fidelity-preserving truncation protocol. (a) Choose two sites, $s_{i}$ and $s_{i+1}$ whose corresponding 
$\chi_{i}$, the shared virtual index, needs to be truncated. Fix the orthogonality center on one of the above 
sites. (b) Perform a contraction over $\chi_{i}$. (c) Further perform an SVD on the above with a cutoff, 
$\Lambda$ resulting in (d) where the corresponding singular values are truncated and renormalized in the $\chi$-subspace (the 
approximation arises out of the truncation). (e) Retrieving the LPDO form by merging the singular values into either 
the left or right tensor depending on the orthogonality center.}
\label{fig:opt_scheme}
\end{figure}

\begin{figure*}
\begin{center}
\begin{tabular}{cp{0.01mm}cp{0.01mm}c}
\subfig{(a)}{\includegraphics[width=0.29\linewidth]{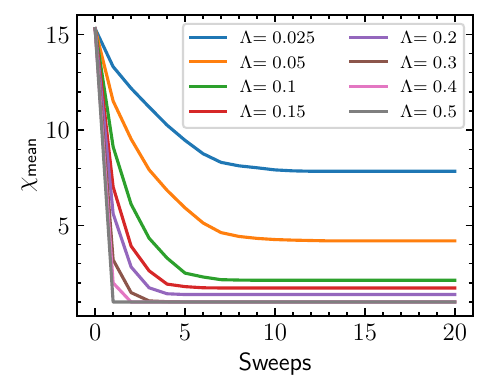}}
&&
\subfig{(b)}{\includegraphics[width=0.29\linewidth]{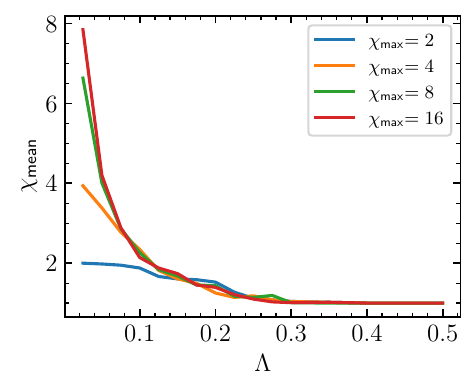}}
&&
\subfig{(c)}{\includegraphics[width=0.29\linewidth]{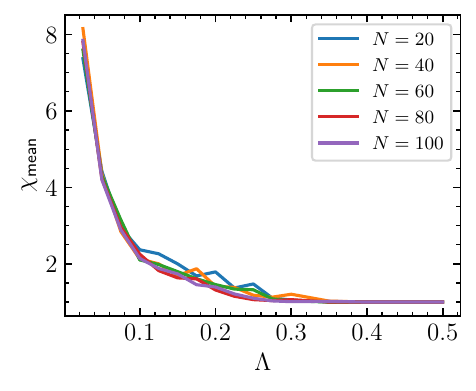}}
\end{tabular}
\end{center}
\caption{(a) To regauge into optimal representation we sweep multiple times across the LPMM$\rho$, with each sweep involving serial two-body updates. Averaged bond dimension, $\chi_\text{mean}$, as a function of sweeps for different cutoff, $\Lambda$ for an initial sub-optimal LPMM$\rho$ of size $N=100$ and $\chi_{\text{max}}=16$. We note that $\chi_{\text{mean}}$ flattens out after a few number of sweeps for a given cutoff as the distribution associated with the singular values at each $\chi$ becomes fixed. Scaling of $\chi_{\text{mean}}$ post fidelity-preserving truncation as a function of the cutoff, $\Lambda$ for (b) different $\chi_{\text{max}}$ with fixed $N=100$, (c) different system sizes $N$ for fixed $\chi_{\text{max}}=16$, with number of sweeps fixed at 20. We observe exponential scaling of $\chi_{\text{mean}}$ as a a function (a) sweeps, and (b, c) cutoff $\Lambda$. To quantify the scaling, in App.~\ref{sec:scaling_at}, we perform a fit to an exponential function, $f(x) = \alpha + \beta e^{-\gamma x}$ and obtain the fit coefficients, $\alpha, 
\beta, \gamma$ as in Fig.~\ref{fig:at_fit_coeffs}.}
\label{fig:at_chi_avg_vs_cutoff}
\end{figure*}

The application of local noisy quantum channels $\mathcal{E}$ to tensor network representations of density operators is illustrated in Fig.~\ref{fig:noise_cartoon_sub_opt}(b) (for a more detailed description, see Fig.~\ref{fig:lpdo_channel} in App.~\ref{appa:channel_application}). This time dynamics is forward integrated by contracting the $\mathcal{E}, A$ and $A^\dagger$ tensors together. The composite tensor is then reshaped into an $s\chi^2 \times s\chi^2$ dimensional matrix, and
a truncated SVD is performed. This results in a new internal index $\kappa_{i}$-index~\cite{Verstraete2004, Werner2016, Jamadagni2024} (illustrated in Fig.~\ref{fig:noise_cartoon_sub_opt}(c), and returns the tensors to the LPDO canonical form. The
singular values of the truncation represent the local mixture probabilities and operationally, truncation 
implies keeping $\kappa$ singular values that equal the dimension of the index. 

This truncation reduces the mixture probabilities' $L_{1}$-norm and we must therefore renormalize~\cite{Werner2016} the probabilities. Take $\mathcal{N}_{\kappa,i} = \sum_{j=1}^{j_{max}}p_{j,i}$ as the $L_1$ norm of the $j$ probabilities at a site $i$.  While $\mathcal{N}_{\kappa,i}=1$ ideally, we use $\mathcal{N}_{\kappa,i}$ to eliminate machine precision errors. The probability lost is $\delta = \sum_{p_{j,i}<\Lambda}p_{j,i}$. Hence, each probability is renormalized as $p_{j,i} \rightarrow p_{j,i}/(\mathcal{N}_{\kappa,i} - \delta)$.

Similarly, the $\chi$-space is renormalized by a similar prescription, but this time using the $L_2$ norm for the $c_{i}$ coherent amplitudes, rather than the $L_1$ norm~\cite{Jamadagni2024} for probabilities. The total norm is $\mathcal{N}_{\chi_i} = \big(\sum_{j=1}^{j_{max}}c_{j}^{2}\big)^{1/2}$ and, keeping $\chi \leq j_{max}$ singular values, the truncated norm is $\delta_{\chi} = {\big(\sum_{j=\chi+1}^{j_{max}}c_{j}^{2}}\big)^{1/2}$, leading to a renormalization $c_i \rightarrow c_i/\sqrt{\mathcal{N}_{\chi_i}^2 - \delta_{\chi}^{2}}$. Since the noise-depolarized state is separable, we originally expected that truncation with low cutoff 
(on the order of $10^{-8}$) in the virtual entanglement space would lower the bond dimension to $\chi=1$. However, we observed that this was not the case. Given that the fidelity with the ideal LPMM$\rho$, a known separable state lacking entanglement, was unity we conclude that the noise LPMM$\rho$ is sub-optimal and conclude that $\chi$ should be pruned. 
These steps are illustrated in Fig.~\ref{fig:opt_scheme}.

To optimize the sub-optimal LPMM$\rho$, we first describe the generalized notion of \textit{fidelity-preserving} truncation. That is, for typical MPS, the infidelity (or approximation error) between the original and the truncated MPS scales linearly with the cutoff, $\delta_\chi$. However, for the sub-optimal LPMM$\rho$ we find that this linear scaling is not generically the case. The intuition is, and empirical results presented will show, that, for the separable LPMM$\rho$, the approximation error (using fidelity as a metric) scaling is more forgiving. We therefore contract nearest neighbor tensors and again perform a SVD but now with a higher value for cutoff, $\Lambda$ followed by a renormalization in the $\chi$-subspace as described above. 

Further, to re-gauge into the optimal representation, we repeat the update across neighboring sites in series and sweep multiple times across the chain for convergence, see Fig.~\ref{fig:at_chi_avg_vs_cutoff}(a). Intuitively, the spurious representation-theoretic entanglement (meaning $\chi>1$ while the state is not entangled as evinced by unit fidelity with the optimal LPMM$\rho$) suggests one may select a large cutoff value to brute-force prune $\chi$. As illustrated by the reduction in $\chi$ in Figs.~\ref{fig:at_chi_avg_vs_cutoff}(b), (c), this is indeed the case. One might be tempted to assume, guided by MPS intution, that choosing a higher cutoff would reduce the fidelity of the LPMM$\rho$ obtained by heavy truncation. On the contrary, Figs.~\ref{fig:at_chi_fid_norm}(a), (b), show that the numerically computed fidelity and norm of the final LPMM$\rho$ are retained despite the large truncation parameter. 


Figure~\ref{fig:at_chi_avg_vs_cutoff}(b, c) highlights how a lower cutoff bounds the reduction in $\chi$.  That is,
choosing lower values for the cutoff, bounds the truncation in $\chi$. To reason this behavior, we note that each 
SVD truncation followed by renormalization, redistributes the associated singular values. In the case that we choose
a lower cutoff, the redistributed singular values may no longer remain sensitive to the lower cutoff leading to their
distribution becoming fixed beyond a certain number of sweeps. In Fig.~\ref{fig:at_chi_avg_vs_cutoff}(a), we 
see that $\chi_{\text{mean}}$ flattens out after few sweeps implying the distribution of the singular values at 
each $\chi$ in the chain becomes fixed and thereby remains insensitive to the cutoff. Therefore, a more nuanced approach 
to force a change in the distribution of the singular values is to exploit the isometries in the mixture subspace 
to influence the coherent correlations. To this extent, in the next section we present a Riemannian manifold based
optimization technique that employs operations in the $\kappa$-subspace to effectively prune $\chi$.


\subsection{Riemannian Optimization}
\label{sec:Riemann}

The virtual mixture space of the LPDO is endowed with an isometric\footnote{A matrix $V$ is said to a isometry if it satisfies $V^{\dag}V=\mathds{1}$.} gauge freedom. That is, the action of isometries on internal, virtual degrees of freedom leave the density operator's physical properties invariant. It is therefore possible to representationally optimize the LPDO via the action of the isometries in the $\kappa$-space. In the context of this work, we apply this fact to analyze how the spurious $\chi$ correlations can be pruned. Pruning $\chi$ can thus be considered as an optimization problem wherein we search the space of isometries that reduce some objective function of the information in $\chi$ subspace. As the space of isometries constitute the Stiefel manifold, 
Riemannian manifold-based optimization techniques can be deployed. To this end, we build on the techniques outlined in recent works that deploy these optimization routines in the context of circuit compilation~\cite{Le2025}, state preparation~\cite{Rogerson2024} and Lindbladian dynamics~\cite{Ramirez2024}.

\begin{figure}[t]
    \centering
    \includegraphics[width=0.5\linewidth]{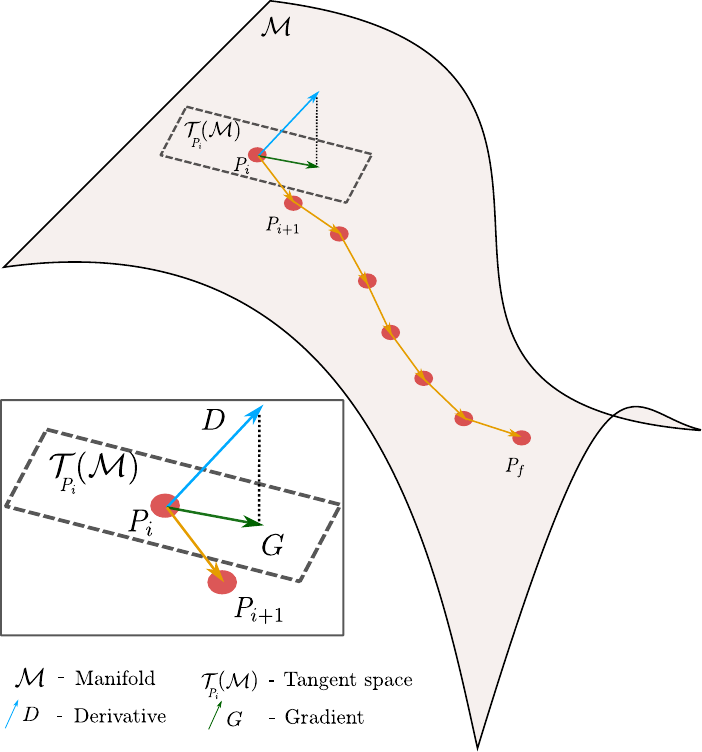}
    \caption{Riemannian manifold based optimization protocol. We initialize the optimization by choosing a random isometry, labeled as $P_{i}$, on the Stiefel manifold, $\mathcal{M}$. To generate the next update in the manifold, $P_{i+1}$, we compute the numerical derivative of the objective function, $D$ with respect to $P_{i}$.  We then estimate the gradient, $G$ by projecting $D$ onto the tangent space of $P_{i}$, $\mathcal{T}_{P_{i}}(\mathcal{M})$. Finally, we perform a retraction of the gradient onto the manifold to obtain $P_{i+1}$. In each iteration the optimizer performs a search on the manifold that minimizes the value of an entropic objective function.}
    \label{fig:rie_opt}
\end{figure}

Prior to adopting these techniques to the current scenario, we briefly review Riemannian manifold-based optimization as outlined in Ref.~\onlinecite{Hauru2021} and illustrated in Fig.~\ref{fig:rie_opt}. Consider an objective function $f$ parameterized by an isometry $V$.
We begin the optimization by randomly choosing an initial element from the Stiefel manifold denoted by $P_{i}$. In each iteration, we compute the derivative of the objective function with respect to the isometry, $D = \partial{f}/\partial{V}$ (as indicated by the blue arrow). Next, we compute the gradient $G$ (green arrow) by projecting the derivative into the tangent space $\mathcal{T}_{V}(\mathcal{M})$ associated with the isometry $V$. As the gradient transports one off the manifold, we then perform a retraction of the gradient back onto the manifold (yellow arrow)~\footnote{In addition to the gradients, optimization routines often require vector transport protocols to estimate the direction of search.}. As a result the isometry $P_{i}$ is updated to $P_{i+1}$.
While there are several strategies to compute the derivative of the objective function with respect to an 
isometry~\cite{Evenbly2009}, we resort to numerical differentiation of the objective function, following 
Ref.~\citenum{Rogerson2024} using Zygote.jl~\cite{Zygote2018}. Further, given the well-defined cost function 
and its gradient (obtained by the projection of the differential), we perform the optimization using the trust-regions solver from 
Manopt.jl~\cite{Bergmann2022} in the Stiefel isometry manifold as in Manifolds.jl~\cite{Axen2023}. 

Having sketched Riemannian optimization, we now discuss the objective functions to be optimized. From an implementation perspective, we note that it is important for the objective function to return a scalar value as this would facilitate the computation of numerical derivatives. Given this constraint, we select two entropy-based objective functions.

\begin{figure*}
\begin{center}
\begin{tabular}{cp{0.01mm}c}
\subfig{(a)}{\includegraphics[height=6.5cm, keepaspectratio]{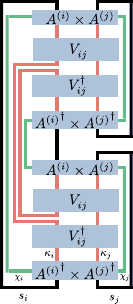}}
&&
\subfig{(b)}{\includegraphics[height=6.5cm, keepaspectratio]{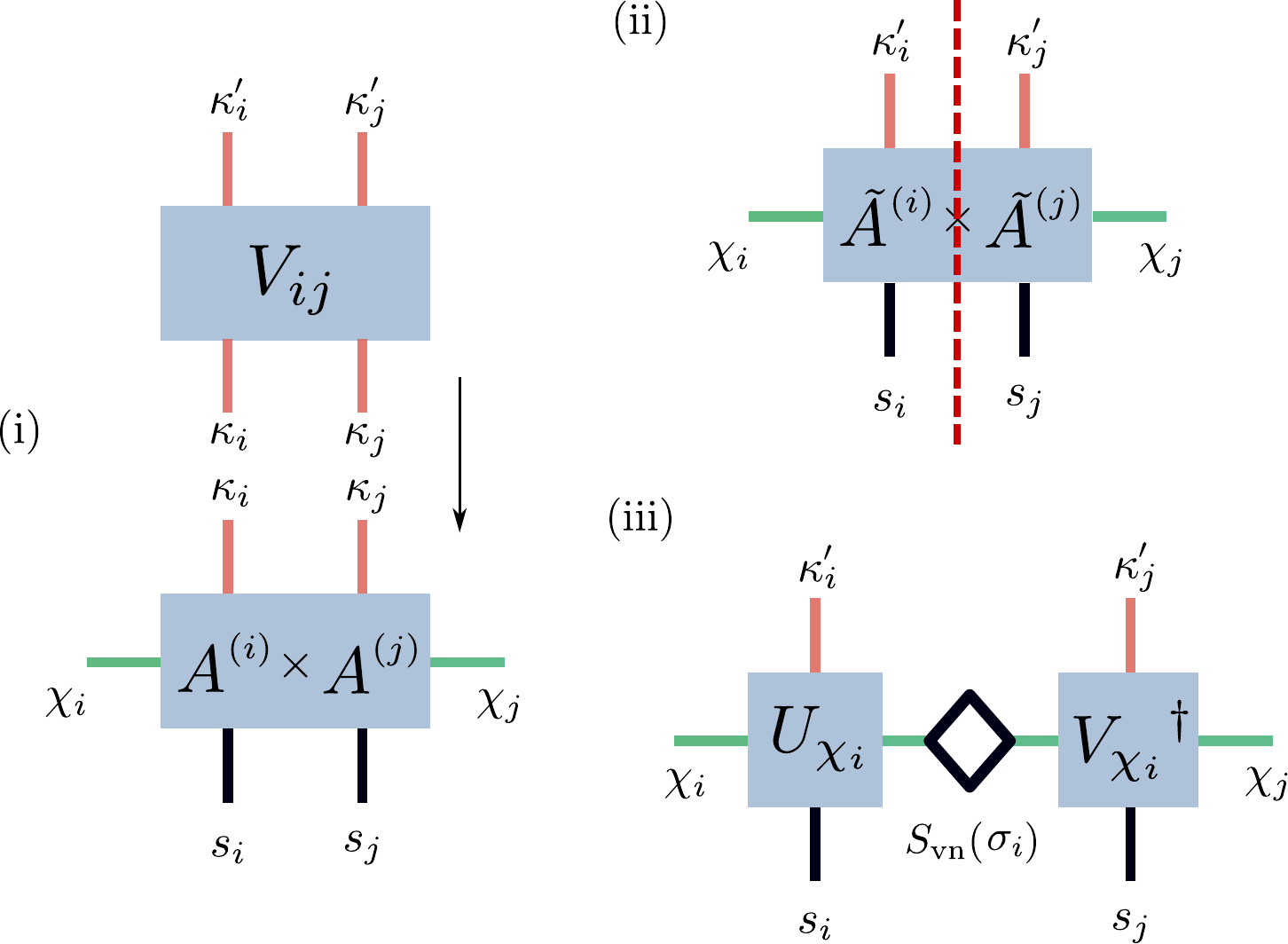}}
\end{tabular}
\end{center}
\caption{Entropy based objective functionals. The Riemannian optimization protocol involves 
optimizing for an isometry, $V_{ij}$ on the Stiefel manifold in the $\kappa$-subspace that minimizes the 
(a) second Renyi entropy, $S_{\text{sr}}$, obtained by tracing the square of the two-site reduced density 
matrix, (b) Von-Neumann entropy, $S_{\text{vn}}$, computed from the singular values, $\sigma_{i}$'s, obtained by 
performing an SVD on the updated two-site tensor.}
\label{fig:obj_funcs}
\end{figure*}

\begin{figure*}[t]
\begin{center}
\begin{tabular}{cp{0.01mm}c}
\subfig{(a)}{\includegraphics[width=0.4\linewidth]{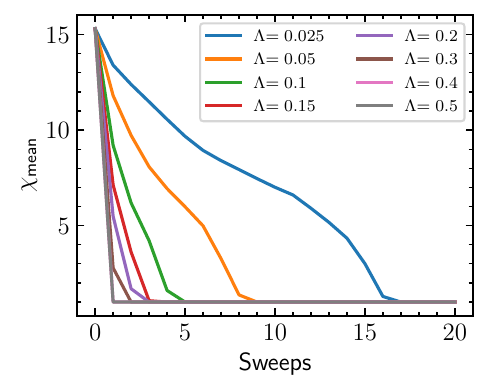}}
&&
\subfig{(b)}{\includegraphics[width=0.4\linewidth]{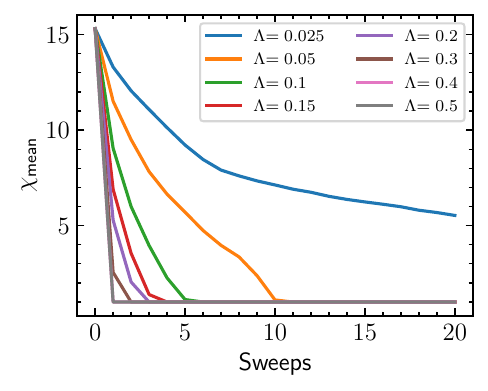}}
\end{tabular}
\end{center}
\caption{Averaged bond dimension across the chain, $\chi_{\text{mean}}$, as a function of sweeps for different 
cutoff $\Lambda_{c}$ using the objective functions (a) second Renyi entropy, $S_{\text{sr}}$, (b) Von-Neumann entropy, 
$S_{\text{vn}}$. We notice that the objective function based on $S_{\text{sr}}$ minimization converges faster in comparison
to the objective function based on $S_{\text{vn}}$. We also note that $S_{\text{sr}}$ minimization 
is more efficient computationally in comparison to the $S_{\text{vn}}$ minimization as the latter involves the extraction 
of singular values thereby leading to an additional SVD operation.}
\label{fig:at_chi_avg_vs_sweeps_opt}
\end{figure*}

The second Renyi (sr) entropy $S_{\text{sr}}$, given by 
\begin{equation}
S_{\text{sr}} = -\log\text{Tr}(\rho^{2}),
\end{equation}
is an entropic measure for a density operator. In our case $\rho = \rho_{\text{2-RDM}}$ is the reduced density matrix of the 
two sites. As in the MPS, we construct the reduced density matrix by tracing over the tensors outside our region of interest. 
Then, in the context of the LPDO, Ref.~\citenum{Hauschild2018} showed how a counter-intuitive contraction sequence (see 
Fig.~\ref{fig:obj_funcs}(a)), is used to compute $\rho^2$ and in turn to obtain the second Renyi entropy, $S_{\text{sr}}$. If 
this contraction sequence is not used, since $V^{\dag}V=\mathds{1}$ the isometries would have no effect on this physical 
observable. In other words, the introduced measure, $S_{\text{sr}}$ computed using the specific contraction sequence is a signature of representational entropy i.e., the entropy of representation of LPMM$\rho$.

\begin{algorithm}[H] 
\begin{algorithmic}[1]
    \FOR{$\text{sweep}=1$ to $M$}
        \FOR{$i=1$ to $N-1$}
            \STATE Contract over $\chi_{i}$ 
            \STATE Invoke optimizer: Minimize $S_\text{sr}$ or $S_{\text{vn}}$ by searching the space of isometries in
            the Stiefel manifold, see Fig.~\ref{fig:obj_funcs}
            \IF{Optimizer converges} 
            \RETURN {Optimized isometry $V_{ij}$} 
            \ELSIF{Terminate after $n_{\text{iter}}$ iterations}
            \RETURN {Isometry $V_{ij}$} 
            \ENDIF
            \STATE Update the initial $\chi_{i}$ contracted tensor by the action of the isometry, $V_{ij}$
            \STATE Perform truncation and renormalize by choosing a higher value for the cutoff, $\Lambda$ (SVD)
            \STATE Update the two sites of the LPDO, $\Phi \rightarrow \tilde{\Phi}$ 
        \ENDFOR
    \ENDFOR  
\end{algorithmic}
\caption{Optimization routine for pruning $\chi$}
\label{alg:opt_routine}
\end{algorithm}

Another signature, encoding the representational entropy, is the Von-Neumann (vn) entanglement entropy, $S_{\text{vn}}$ associated with the singular value $\sigma_{i}$'s at the bond $\chi$. These values are obtained by performing a SVD at the corresponding bond with $S_{\text{vn}}$ given by~\cite{Kevin2021}
\begin{equation}
\label{eq:ent}
S_{\text{vn}} = -\sum_{i}\lambda_{i}\log\lambda_{i}; \quad \lambda_{i} = \frac{\sigma_{i}^2}{\sum_{i}\sigma_{i}^2}.
\end{equation}
The central theme in the optimization routine is to find isometries in the mixture subspace that disentangle the sub-optimal LPMM$\rho$. In other words, we minimize the entropy-based objective functions by searching for suitable \textit{disentangling isometries} in the Stiefel manifold acting in the $\kappa$ subspace. 
Given a sub-optimal LPMM$\rho$, say $\Phi$ consisting of $N$ sites, we present the steps involved in the optimization protocol in Alg.~\ref{alg:opt_routine}. 
To prune $\chi_{i}$, we contract over $\chi_{i}$ and then invoke the 
optimization solver that explores the space of isometries whose dimension is given by $\kappa_{c} \times \kappa_{c}$,
where $\kappa_{c} = \kappa_{i} \times \kappa_{i+1}$. 
We observe that in the process of lowering the entropies, the optimizer finds isometries that redistribute the spectral weight onto the leading singular values (making them the heavy weight entries). However this process may introduce additional lower weight entries (contributing negligibly to the entropy) which increase $\chi$. This elongation of the tail of the singular values was also noted in earlier studies, for instance in Ref.~\onlinecite{Hauschild2018}. The increase in the tail is indicative that choosing a very low cutoff (on the order of $10^{-8}$ as in a conventional MPS setting) in the truncation, after the isometric update (step 11 in Tab.~\ref{alg:opt_routine}) might stall the optimization process due to the growth in $\chi$.

Therefore, we invoke the fidelity-preserving truncation protocol, as discussed in Sec.~\ref{sec:fpt}, and choose a higher 
value for the cutoff, $\Lambda$ post the termination of the optimizer. Figure~\ref{fig:at_chi_avg_vs_sweeps_opt} illustrates 
that the optimization routine remains an effective tool in pruning $\chi$ to 1. We briefly discuss the different cutoff 
regimes and their impact on pruning $\chi$ using the two numerical methods. For low cutoff regimes, the fidelity-preserving 
truncation protocol can be made more effective by integrating them with optimization protocols. However, in the high cutoff
regimes, the role of the optimizer is minimal as the truncation after few sweeps already prunes $\chi$ to 1.

To validate the protocol, in Fig.~\ref{fig:at_chi_fid_norm}(c, e), we present the fidelity of the Riemannian optimized state with respect to the initial state, as well as the norm (d, f) of the Riemannian optimized representation. We observe that, while the norm is 
conserved, the infidelity measure for $\Lambda \leq 0.3$ is 3 to 4 
orders of magnitude greater than the infidelity obtained using the fidelity-preserving truncation protocol, Fig.~\ref{fig:at_chi_fid_norm} (a). Furthermore, we note 
that the infidelity measures obtained using the objective function based on the Von-Neumann entropy 
$S_{\text{vn}}$ are 1 to 2 orders of magnitude lower than those based on the second Renyi entropy, $S_{\text{sr}}$, see 
App.~\ref{sec:at_validation} for more details.

\subsection{Symmetry and Injectivity}
In this section, we shift from numerical methods to present complementary analytical tools to tame the $\chi$-complexity. A difference from the previous analysis is that, instead of mapping a suboptimal LPMM$\rho$ to its optimal counterpart, we shall first discuss how unitaries induce spurious correlations into the optimal LPMM$\rho$. We will then present how isometries in the virtual mixture space prune the unitarily-induced spurious entanglement. 

We begin by noting that the action of two-qubit or higher order unitary on the optimal LPMM$\rho$ induces spurious entanglement.
That is, the application of a unitary on the physical indices increases $\chi$, see Fig.~\ref{fig:lpdo_uni}(b) in App.~\ref{appa:unitary_application}. 
The unitary action leaves a maximally mixed state invariant, however representationally this turns an optimal LPMM$\rho$ into a sub-optimal one
due to the increase in $\chi$. To establish the existence of a $\kappa$-virtual space isometry via injectivity that effectively inverts the unitary action i.e., prunes $\chi$, we first review the definition of weak symmetry followed by the notion of weak injectivity and the correspondence between the both as discussed in Ref.~\citenum{Guo2025}.

A density matrix $\rho$ is weakly symmetric under $U$ if it is invariant under the conjugation by $U$ i.e., $U\rho U^{\dag} = \rho$. Strong symmetry is the stricter condition that $ U\rho = e^{-i\phi} \rho$ which also implies weak symmetry. The maximally mixed density operator, $\rho_{\mathds{1}}$ as introduced in Eq.~\ref{eq:rho_MM}, is weakly symmetric under \textit{any} unitary operator $U$ since  $U\rho_{\mathds{1}}U^{\dag} = \rho_{\mathds{1}}$. It is however, not strongly symmetric under the left or right action of $U$. 

Next, we introduce the notion of weak injectivity. In the context of the LPDO, weak injectivity can be defined as a map that leaves the local tensors $A^{(i)}$ as injective i.e., see Eq.~\ref{fig:lpdo_injectivity} as introduced in Ref.~\citenum{Guo2025}. 
This injectivity relation states that a unitary, acting on the physical indices, is equivalent to other isomteries acting on both types of virtual indices, modulo a phase. Alternatively, the action by the physical unitary and the inverse of the virtual isometries leaves the tensor invariant. 

\begin{equation}
\begin{split}
\includegraphics[width=0.6\linewidth]{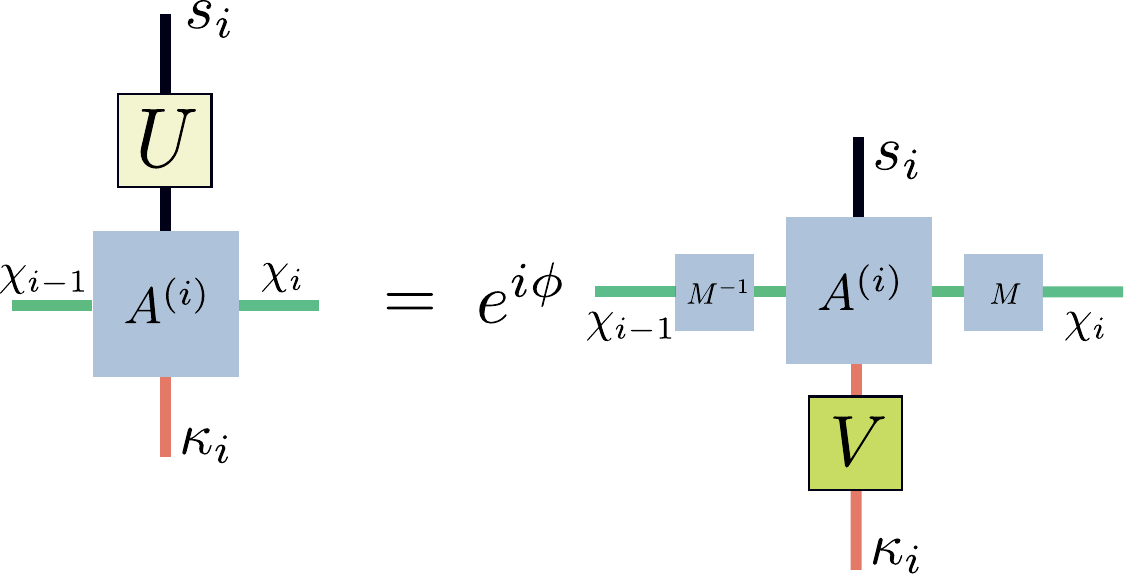} & \\
 \sum_{s'_{i}}U_{s_{i}}^{s'_{i}}{[A^{(i)}]}_{s'_{i}}^{\chi_{i-1}, \chi_{i},\kappa_{i}} =   \textcolor{white}{[M]_{\chi_{i-1}}^{\chi_{i-1}'}V_{\kappa_{i}}^{\kappa'_{i}}{[A^{(i)}]}^{s_{i}}_{\chi'_{i-1}, \chi'_{i}, \kappa'_{i}}} \\
    e^{i\phi}\sum_{\chi'_{i-1}, \chi'_{i}, \kappa'_{i}}[M^{-1}]_{\chi_{i-1}}^{\chi_{i-1}'}[M]_{\chi_{i}}^{\chi_{i}'}[V]_{\kappa_{i}}^{\kappa'_{i}}{[A^{(i)}]}^{s_{i}}_{\chi'_{i-1}, \chi'_{i}, 
    \kappa'_{i}}
\end{split}
\label{fig:lpdo_injectivity}
\end{equation}


In Ref.~\citenum{Guo2025}, the authors establish the existence of a weakly injective map for weakly symmetric LPDOs. In the current scenario, it implies that the action of some unitary, $U$ acting on the physical indices and inducing spurious $\chi$, is equivalent to the action of some isometry $V$ in the $\kappa$-subspace and isometries in the $\chi$ space. However, in this case the isometry in the $\chi$-space is the identity , up to a $-1$ phase factor, so we ignore it in our discussion. Therefore, the spurious growth of $\chi$ in the coherent space can be pruned by the action of $V^{\dag}$ in the $\kappa$-subspace. Thus, by invoking the existence of injective maps, we establish the fact that the unitary action on the optimal LPMM$\rho$, inducing fictitious $\chi$, can be negated by the action of some isometry in the $\kappa$-subspace. 

However, the weakly injective property does not provide any constructive numerical method to explicitly compute the isometry which corresponds to a given unitary. In the case of the LPMM$\rho$, we present two different strategies to compute the isometry in the mixture subspace. We begin by considering the optimal LPMM$\rho$ wherein each $\kappa_{i}=2$ with $\chi=1$.  One strategy is based on analytic arguments while another highlights how isometric optimization can be used as a tool to identify the injective isometry.

The action of unitary, $U$ on one the LPPM$\rho$'s tensors, $A^{(i)}$, is given by 
$UA^{(i)}$. By invoking the definition of $A^{(i)}$ as in Eq.~\ref{eq:A_j}, we have
the following
\begin{equation}
\begin{split}
UA^{(i)} &= U\frac{\mathds{1}}{\sqrt{2}} = \frac{\mathds{1}}{\sqrt{2}}U \\
       &= e^{i\phi} A^{(i)}  V.
\end{split}
\end{equation}
Therefore, from the above we observe that the equivalent isometry $V$ is given by $V=e^{-i\phi}U$, where $\phi$ is a phase. Therefore, $V^{\dag}U = e^{i\phi}\mathds{1}$ up to some global phase, i.e., the action of unitary can be
inverted by the application of $V^{\dag}$ in the $\kappa$-subspace. Similarly, we can explicitly derive the equivalent isometry upto some global phase in the case where we apply two-qubit or higher order $m$-qubit unitaries, say $\tilde{U}$ in the physical space as follows
\begin{equation}
\begin{split}
\tilde{U}\big(A^{(j)} \otimes \cdots \otimes A^{(m)}\big) &= \tilde{U}\big(\frac{\mathds{1}}{\sqrt{2}} \otimes \cdots \otimes \frac{\mathds{1}}{\sqrt{2}} \big) \\ 
       &= \big(\frac{\mathds{1}}{\sqrt{2}} \otimes \cdots \otimes \frac{\mathds{1}}{\sqrt{2}} \big)\tilde{U} \\
       &= e^{i\tilde{\phi}}\big(A^{(j)} \otimes \cdots \otimes A^{(m)}\big)\tilde{V}.
\end{split}
\end{equation}
Therefore, $\tilde{V} = e^{-i\tilde{\phi}}\tilde{U}$, further leading to $\tilde{V}^{\dag}\tilde{U}=e^{i\tilde{\phi}}\mathds{1}$ and is thereby the disentangling isometry in the $\kappa$ subspace inverting the action of $\tilde{U}$ in the physical space. 


On the other hand, invoking the optimization routines as introduced in Sec.~\ref{sec:Riemann} is also a route towards identifying the injective isometries $\tilde{V}^{\dag}$. Since these isometries return the LPMM$\rho$ to its original form (with $\chi=1$) they effectively disentangle, or in other words prune, the spurious entanglement induced by $\tilde{U}$. In this case the action of the isometry $\tilde{V}$ is equivalent to $\tilde{U}$ up to some $SU(2)$ unitaries on the individual sites. In practical calculations the tensor network representation is also sub-optimal in the virtual mixtures space, with $\kappa_i \ge 2$. In this case, the optimization routines provide an alternative to construct the the disentangler $\tilde{V}^{\dag}$, modulo some gauge $SU(\kappa_{i})$ freedoms acting on the individual sites. In future research it will be of interest to further investigate these properties and automate the construction of more general injective maps. 



\section{Gauging states away from the LPMM$\rho$\label{sec:away_from_mms}}
Moving away from the completely depolarized limit, a few previous studies have optimized the representation of mixed states away from the maximally mixed state. In Ref.~\onlinecite{Nguyen2018}, the authors prepare mixed states away from the maximally mixed state by combining real time evolution with unitary evolution in the $\kappa$-subspace. The unitary evolution results in the optimization of representational entropy associated with the $\chi$-virtual bonds. In a more recent work, in Ref.~\onlinecite{Mueller2024}, a similar optimization scheme is employed to prune the representation. The former of the above works employed a conventional low truncation value while the latter briefly described the impact on representation when choosing higher truncation values.  This leads us to an important and unexplored scenario of how truncation influences the representation of mixed states that are away from LPMM$\rho$, which we explore in this section. We use the infidelity measure given by $\mathcal{I}_{\sigma} =  1-\mathcal{F}(\sigma, \rho_{\mathds{1}})$ as a metric to quantify the divergence of the mixed state $\sigma$ from the maximally mixed state, $\rho_{\mathds{1}}$.

Let us now generalize the results of the prior section by analyzing the impact of truncation and optimization routines on partially mixed states, i.e., $\rho_{0}$ satisfying $\mathcal{I}_{\rho_{0}} = \Delta$ for some given $\Delta$. The prior sections were limited to $\Delta=0$. To generate such a $\rho_{0}$, we choose the rates $\gamma_{d}$ and $\gamma_{b}$, the dephasing and bitflip rates respectively, to be smaller than 0.5 (the fully depolarized limit). Further, denote the state obtained after the truncation cutoff $\Lambda$ and optimization, with respect to an input state $\rho_{0}$, as $\rho_\text{opt}^{\Lambda}$. 
Let us first summarize the main result of the section, which we demonstrate next, as follows: Given a mixed state $\rho_{0}$, satisfying $\mathcal{I}_{\rho_{0}} = \Delta$ for some given $\Delta$ and some given algorithmic precision $\Omega$, i.e. the infidelity between the initial and the optimized representation, the truncation/optimization routines guarantee a truncation $\Lambda$ such that $1-\mathcal{F}(\rho_{0}, \rho_{\text{opt}}^{\Lambda}) = \Omega$. For the case of $\Delta=1$, $\rho_{0}$ is pure and can be represented as an MPS. In this case, the $\Lambda$ truncation causes an approximation error ($\varepsilon = \sum_{i\leq \Lambda}\sigma_i^2$) which in this limit coincides with the infidelity, $\Omega$, between the original and truncated states. In a more generic case, where $\Delta < 1$, the truncation $\Lambda$ is more forgiving. Specifically, as shown below, it linearly interpolates between the well established MPS precision errors limit and the fidelity preserving truncation highlighted in Sec.~\ref{sec:fpt}.

\begin{figure*}[t]
\begin{center}
\includegraphics[width=0.9\linewidth]{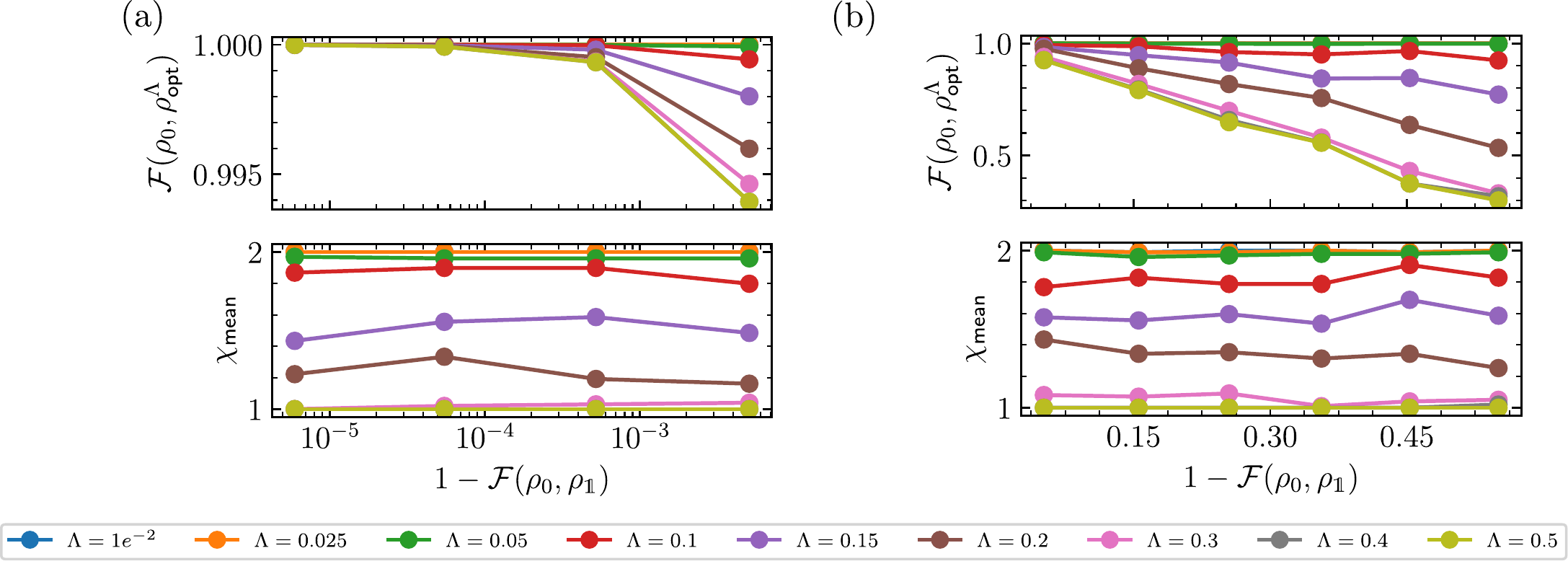}
\end{center}
\caption{The effect of $\Delta$-adaptive truncation on mixed states of size $N=100$ characterized by $\chi_\text{max}=2$. (Top) The impact of truncation quantified by drift in fidelity, $\mathcal{F}(\rho_{0}, \rho_\text{opt}^{\Lambda})$, of the mixed state obtained after truncation $\rho_\text{opt}^{\Lambda}$ with respect to the initial state $\rho_{0}$, for different states $\rho_{0}$ satisfying $\mathcal{I}_{\rho_{0}} = \Delta$ for $\Delta \in$  (a) [$10^{-6}, 10^{-2}$], (b) $[10^{-2}$, 0.6]. (Bottom) The average of the virtual bond dimensions in the final state $\rho_\text{opt}^{\Lambda}$, $\chi_{\text{mean}}$ as a function of mixed states $\rho_{0}$ satisfying $\mathcal{I}_{\rho_{0}} = \Delta$. In (a) for states relatively close to the maximally mixed state, the impact of choosing a high truncation retains the fidelity upto two orders while effectively compressing the representation. A lower value for truncation retains fidelity however does not compress the representation as $\chi_\text{mean}$ remains as in the initial state. In (b), as we move relatively far from the maximally mixed state, the fidelity drops almost linearly with higher truncation leading to poorer fidelities and high compression leading to a non-faithful representation. As in the earlier case, a lower value for truncation retains fidelity with not much compression in the virtual bond dimension, $\chi$. }
\label{fig:fp_fid_chi_epsilon}
\end{figure*}

\subsection{$\Delta$-adaptive Truncation}
In Sec.~\ref{sec:fpt}, we analyzed the impact of truncation on the maximally mixed state and noticed the entanglement pruning to be fidelity preserving, even at higher values of truncation. However, as we consider mixed states, $\rho_{0}$ that satisfy $\mathcal{I}_{\rho_{0}} = \Delta$ for some $\Delta$, choosing a larger truncation cutoff not only prunes the spurious correlations but also other coherent correlations that represent the state. For instance, in the limit of $\Delta=1$, i.e., the case of the MPS, choosing a huge cutoff for truncation leads to an unfaithful representation. In Fig.~\ref{fig:fp_fid_chi_epsilon}, we present the scaling of $\mathcal{F}(\rho_{0}, \rho_\text{opt}^{\Lambda})$ as we move away from the maximally mixed state. The approximation error, as quantified by the decrease in fidelity, is almost linear as we move away from the maximally mixed state for a given cutoff $\Lambda$. The absolute of the linear slope increases with increasing cutoff implying the fidelity decreases as we increase the cutoff for a given mixed state. We also note that the compression in the state representation, captured by $\chi_\text{mean}$ (the average of the virtual-bonds, $\chi_{i}$) increases with a higher truncation, however is accompanied by decrease in fidelity, $\mathcal{F}(\rho_{0}, \rho_\text{opt}^{\Lambda})$. To summarize, our result interpolates between the conventional controlled MPS approximation scaling in the limit $\Delta=1$ and the fidelity preserving truncation, at $\Delta =0$.  Overall, this result highlights how, for a desired approximation error, the truncation is rescaled by a factor $\Delta^{-1}$.

\begin{figure*}[t]
\begin{center}
\includegraphics[width=0.9\linewidth]{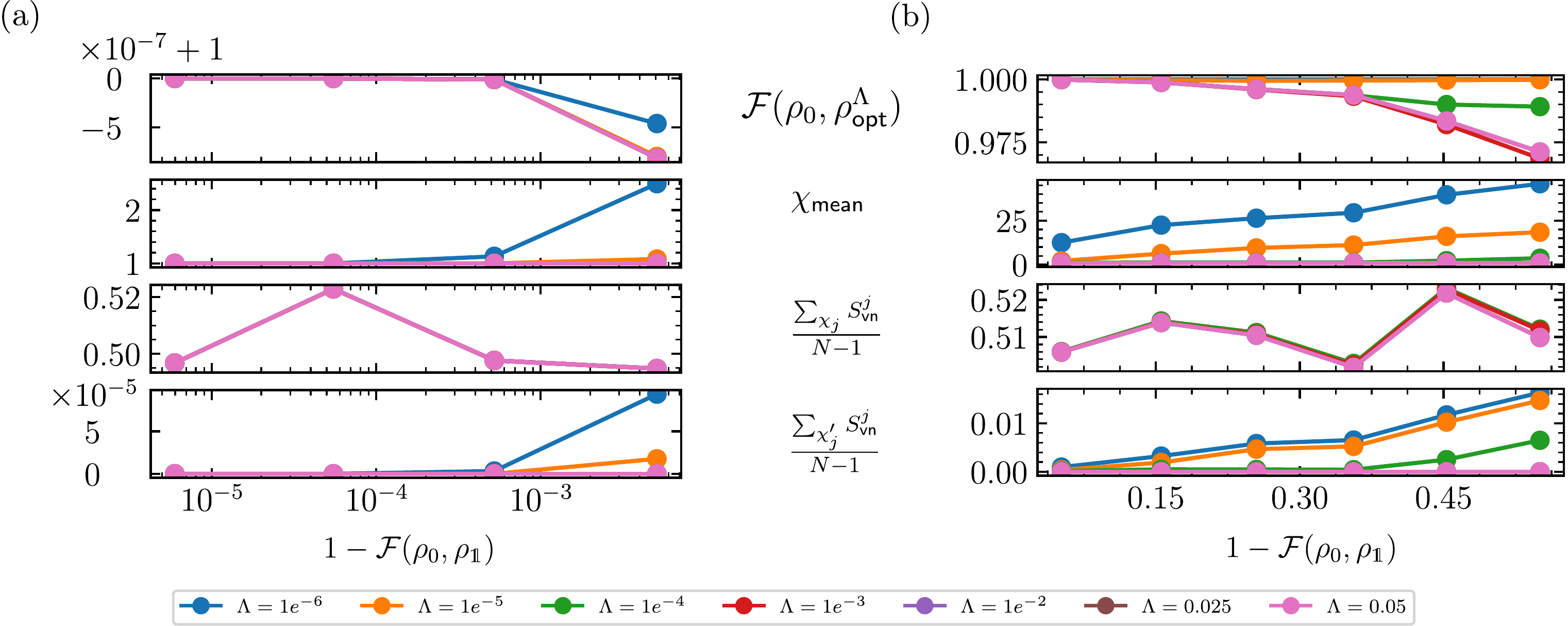}
\end{center}
\caption{Riemannian manifold based optimization to prune correlations in mixed states away from the maximally mixed states of size $N=100$ characterized by $\chi_\text{max} = 2$ . (Top) $\mathcal{F}(\rho_{0}, \rho_\text{opt}^{\Lambda})$ for initial states $\rho_{0}$ satisfying $\mathcal{I}_{\rho_{0}} = \Delta$ for $\Delta \in$ (a) [$10^{-6}$, $10^{-2}$], (b) [$10^{-2}$, 0.6]. (Middle) The average of the bond dimensions in the final state $\rho_\text{opt}^{\Lambda}$ given by $\chi_\text{mean}$ as a function of the mixed states $\rho_{0}$ satisfying $\mathcal{I}_{\rho_{0}} = \Delta$. (Bottom) Average entropy across all the virtual bonds pre and post optimization. We observe that the optimization routine retains fidelity on the order of single float precision for (a) states close to the maximally mixed states while in (b) are relatively higher in comparison to the $\Delta$-adaptive truncation. The compression in the representation, given by $\chi_\text{mean}$ remains effective for higher truncation i.e., choosing a higher truncation reduces the $\chi$. However, for a lower choice of the truncation, we observe a growth in $\chi$. The average entropy measures indicate that the entanglement pruning is effective i.e., the averaged entropy of the optimized state drops significantly and the reduction remains almost consistent for all states away from the maximally mixed state at all truncation values.}
\label{fig:svd_opt_chi_epsilon}
\end{figure*}

\subsection{Riemannian Optimization}
We now study the impact of Riemannian manifold based optimizers on pruning entanglement on mixed states that are away from the maximally mixed state. As earlier, we consider mixed states $\rho_{0}$ that satisfy $\mathcal{I}_{\rho_{0}} = \Delta$, for some given $\Delta$. We compute the fidelity given by $\mathcal{F}(\rho_{0}, \rho_\text{opt}^{\Lambda})$ to track the impact of the truncation while using the optimizer to prune entanglement. In Sec.~\ref{sec:Riemann}, we analyzed the impact of optimizer on pruning the entanglement by choosing two different objective functions. In the following analysis, we minimize the Von-Neumann entropy based objective function, $S_\text{vn}$, as in Eq.~\ref{eq:ent} as we expect similar behavior for the other objective function based on second Renyi entropy, $S_\text{ar}$. Fig.~\ref{fig:svd_opt_chi_epsilon} illustrates how coupling $\Delta$-adaptive truncation catalyzes Riemannian optimization. This allows us to minimize approximation errors and representation theoretic complexity away from the completely depolarized state.
We attribute this to the optimizer shifting the distribution of the singular values i.e., $\sigma'_{0} > \sigma_{0}$ and $\sigma'_{i} < \sigma_{i}$, where $\sigma_{j}$ and $\sigma'_{j}$ are singular values from consecutive optimization sweeps at some virtual bond $\chi$. We quantify the entanglement pruning by computing the average of the Von-Neumann entropy across all the virtual bonds at the beginning (end) of the optimization. It is given by $\frac{\sum_{\chi_{j}}S_\text{vn}^{j}}{N-1}$ ($\frac{\sum_{\chi'_{j}}S_\text{vn}^{j}}{N-1}$), where $\chi_{j}$ ($\chi'_{j}$) represent the bonds at the beginning (end) of the optimization. As we move away from the maximally mixed state, we notice a dip in the fidelity, $\mathcal{F}(\rho_{0}, \rho_\text{opt}^{\Lambda})$, however this dip is no longer linear as in the previous case and has better fidelity retaining properties. One other important observation is that while the effective entropy of the representation is reduced, this does not automatically imply the virtual bond dimension is minimized. In fact, we notice that at lower truncation $\Lambda$, the average bond dimension, $\chi_\text{mean}$ is significantly higher compared to the initial state where as the entropy is minimized. This is in agreement with previous results as in Sec.~\ref{sec:Riemann} and earlier studies~\cite{Hauschild2018}, where reduction in representation entropy is accompanied by growth in the $\chi$-virtual bond.

\section{Conclusion\label{sec:conclusion}}
To summarize, we have introduced three complementary technical tools (fidelity-preserving truncation, Riemannian manifold based optimization, and injectivity) to optimize the representation-complexity of density operators. We tested and provided quantitative scalings for their performance in the controlled, but physically relevant, setting of the maximally mixed density operators. This setting is chosen because it is known that infinite temperature states are separable (meaning ideally $\chi=1$) but minimizing $\chi$ to the expected level was challenging in practice using the conventional toolkit borrowed from MPSs.  

We establish that minimizing $\chi$, by choosing an SVD truncation parameter $\Lambda$ much larger than in MPS calculations, proved to be an effective strategy to prune $\chi$. We quantified the empirical performance of this fidelity-preserving truncation and also provided quantitative relationships between the number of truncation sweeps and how much $\chi$ was pruned. In doing so, we confirmed that the conventional MPS-like small cutoffs (chosen because in the MPS setting the approximation error is proportional to the cutoff) marginally  reduced $\chi$, but not to unity. 

To compliment truncation, we then explored Riemannian-manifold-based optimization techniques. This involved searching for the isometries , acting in the $\kappa$-subspace and residing on a Stiefel manifold, that minimize the tensor networks representation complexity via $\chi$-entropy objective functions. We notice that, in the case that the SVD-truncation parameter chosen, remains relatively low to completely prune $\chi$, 
the optimization routines constituted effective tools to disentangle the tensor-network's representation by reducing $\chi$ to 1. The quantitative emperical behaviour in terms of optimization parameters and objective functions were provided. Here we found that the von Neumann and second-Renyi entropies were both suitable entropic objective functions.

We also established analytical results on existence of isometries in the $\kappa$-subspace that efficiently prune $\chi$ thereby mapping back to the optimal representation. To do so, we turned the workflow on its head by beginning with an optimal representation and then impairing it. We began with an optimal LPMM$\rho$ (with $\chi_i=1$ and $\kappa_i=2$ for all $i$) and noticed that many-body unitaries increased $\chi$. This is surprising since the LPMM$\rho$ $\rho_\mathds{1}\propto \mathds{1}$ it commutes with all operators and is therefore weakly symmetric under all unitaries, i.e., $U\rho_\mathds{1}U^\dagger = \rho_\mathds{1}$. Since weak symmetry has been recently shown to imply weak injectivity in the LPDO~\cite{Guo2025} we have used this property. More specifically, we proved that, for the LPMM$\rho$, the form of the optimal disentangler which is analytically enforced by injectivity is the hermitian conjugate of the many body unitary applied to the physical indices. Lastly we discussed how the isometric distenanglers, found via optimization, were equivalent to the physical unitary up to local $SU(2)$ operations.

Lastly, we quantified our algorithm's performance in the general setting of partially mixed or depolarized states. Given a mixed state $\rho_{0}$, with an infidelity $\Delta$ with respect to the maximally mixed state and for some given precision $\Omega$ (the infidelity between the initial and optimized representations) we numerically demonstrated that there exists an improved truncation cutoff, $\Lambda\geq\Lambda_{MPS}$ such that $1-\mathcal{F}(\rho_{0}, \rho_\text{opt}^{\Lambda}) = \Omega$. We introduced $\Delta$-adaptive truncation, a routine that generalizes the fidelity preserving truncation introduced in the context of LPMM$\rho$. We establish that employing $\Delta$-adaptive truncation results in fidelity $\mathcal{F}(\rho_{0}, \rho_\text{opt}^{\Lambda})$ that linearly interpolates, as a function of different $\rho_{0}$'s that move away from the LPMM$\rho$, between the MPS and the fidelity preserving truncation limits. We observed that at low truncation values, $\mathcal{F}(\rho_{0}, \rho_\text{opt}^{\Lambda})$ remains high with not much of compression in the representation. On the other hand at higher truncation values, the fidelity $\mathcal{F}(\rho_{0}, \rho_\text{opt}^{\Lambda})$ begins to dip with the representation quickly becoming unfaithful. Next, we investigated the impact of Riemannian manifold based optimizers coupled with the $\Delta$-adaptive truncation on the representation of mixed states. Here, we established that the fidelity $\mathcal{F}(\rho_{0}, \rho_\text{opt}^{\Lambda})$ scales better in comparison to standalone $\Delta$-adaptive truncation i.e., we notice the optimized representation retains most of the features of its sub-optimal counterpart, even for mixed states that are considerably away from LPMM$\rho$. We also observe that the average entropy of the optimized representation is reduced, however for low truncation values and large $\Delta$ the compression remains poor due to the growth in the bond dimension. This limit is the most difficult, with additional algorithmic research being required to drive new advances that minimize representation theoretic resources for the modeling and simulation of mixed states that appear in nature and in qubit experiments. 


In conclusion, using complimentary numerical and analytic tools, we have completely solved the re-optimization and entanglement pruning of the LPMM$\rho$. The methods introduced in this work can be used to investigate a variety of mixed states leading to an impact on i) qubit engineering, ii) reducing the classical resource requirements of simulation of open quantum systems (including noisy circuits~\cite{Arute2019}), and increase the reliability of noise-extrapolation-based error-mitigation strategies\cite{Temme2017, Li2017,Thompson2025}. One other interesting direction is to extend the methods introduced to analyze representations of systems with additional symmetries, for instance systems with strong symmetries~\cite{Moharramipour2024}, or in higher dimensional systems.



\begin{acknowledgments}
This manuscript has been authored by UT-Battelle, LLC, under Contract No. DE-AC0500OR22725 with the U.S. Department of Energy. The United States Government retains and the publisher, by accepting the article for Publication, acknowledges that the United States Government retains a non-exclusive, paid-up, irrevocable, worldwide license to publish or reproduce the published form of this manuscript, or allow others to do so, for the United States Government purposes. The Department of Energy will provide public access to these results of federally sponsored research in accordance with the DOE Public Access Plan. A.G. and E.D. are supported by the U.S. Department of Energy, Office of Science, Advanced Scientific Research Program, Early Career Award under contract number ERKJ420.
\end{acknowledgments}

\appendix

\section{Review of LPDOs\label{app:review_lpdo}}
\begin{figure}[t]
\includegraphics[width=0.6\linewidth]{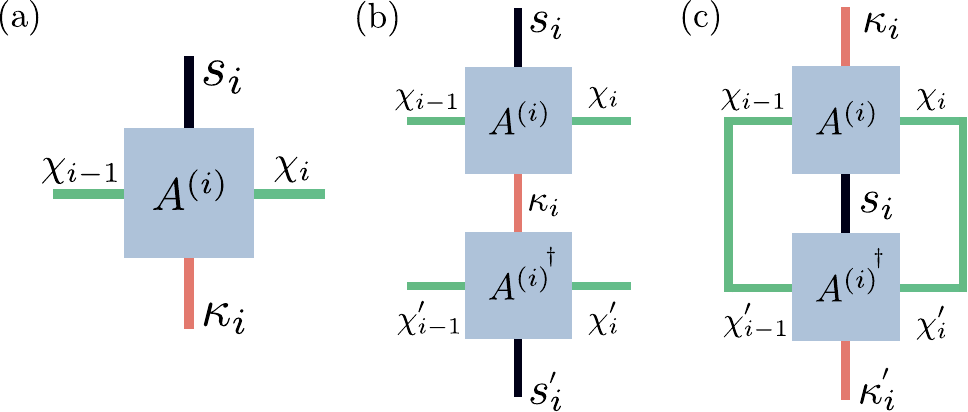}
\caption{(a) Representating an LPDO by decorating each site, $i$ in the MPS with an additional
virtual index, $\kappa_{i}$ the encode the mixture correlations. (b) Contracting over the $\kappa_{i}$
index turning the LPDO to an MPO. (c) Tracing over the physical index, $s_{i}$ in addition to the
coherent $\chi_{i}$ and $\chi_{i-1}$ indices provides access to the probabilities of pure states
that constitute the density matrix.}
\label{fig:lpdo_summary}
\end{figure}

In the following, we briefly review the notion of LPDOs, while also revisiting
some of its features~\cite{Werner2016,Jamadagni2024}. LPDOs are extensions of MPSs,
i.e., we decorate each tensor site in the MPS with an additional index, also 
referred to as $\kappa$-index, that encode the classical mixture correlations, 
see Fig.~\ref{fig:lpdo_summary}(a). Tracing over the $\kappa$-index results in an 
the density matrix i.e., $\rho=A^{(i)}{A^{(i)}}^{\dag}$ (equivalently an MPO), 
see Fig.~\ref{fig:lpdo_summary}(b) and therefore the LPDO representation respects positivity 
by construction. In addition, tracing over the physical index offers access
to pure states that constitute the density matrix as illustrated in Fig.~\ref{fig:lpdo_summary}(c).
In the next sections, we briefly review the application of unitaries 
and noise channels on the LPDO as well present the notion of isometric gauge freedom in the 
$\kappa$-subspace.

\subsection{Unitary application\label{appa:unitary_application}}
\begin{figure}
\includegraphics[width=0.6\linewidth]{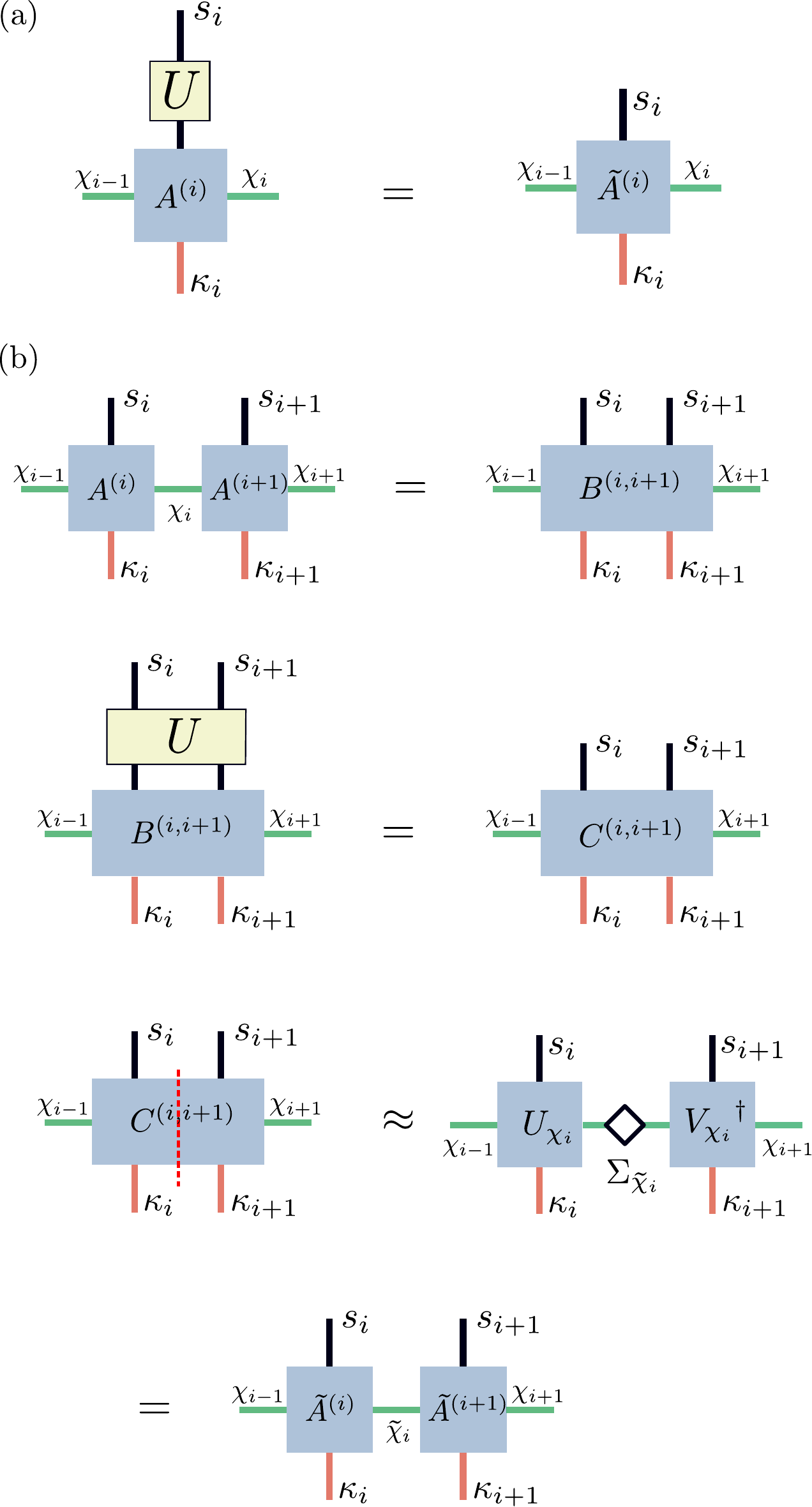}
\caption{Application of (a) single qubit unitary, (b) two-qubit unitary on an LPDO. In (a) the 
application of single qubit unitary updates the site tensor $A^{(i)}$ to $\tilde{A}^{(i)}$ with
$\chi_{i-1}$, $\kappa_{i}$ and $\chi_{i}$ remaining unchanged. In (b) for the application 
of two qubit unitary on sites $s_{i}$ and $s_{i+1}$, we contract over the shared coherent
index, $\chi_{i}$ resulting in the composite object $B^{(i, i+1)}$. We then apply the two-qubit 
unitary resulting in $B^{(i,i+1)}$ being mapped to $C^{(i, i+1)}$. Further, we perform an SVD
on $C^{(i,i+1)}$, followed by truncation and renormalization of the singular values, resulting
in $\chi_{i}$ being updated to $\tilde{\chi}_{i}$. Finally, we return to the LPDO representation 
by integrating the singular values into the left/right tensor depending on where the orthogonality 
center is fixed. }
\label{fig:lpdo_uni}
\end{figure}

We present the application of single and two-qubit unitaries on an LPDO. To this extent, we 
begin by fixing the orthogonality center to the site on which we wish to apply the unitary. In the 
case of two-qubit unitary we choose either one of the sites. The single qubit
unitary application leads to the update in the tensor at site $i$ from $A^{(i)}$ to 
$\tilde{A}^{(i)}$ leaving the mixture and coherent indices untouched, thereby they remain 
the same as before the unitary update, see Fig.~\ref{fig:lpdo_uni}(a). The two-qubit unitary involves 
contracting over the shared coherent index leading to the application of the two-qubit unitary 
followed by performing an SVD on the updated composite tensor thereby returning to the
LPDO representation, see Fig.~\ref{fig:lpdo_uni}(b).

\subsection{Channel application\label{appa:channel_application}}
\begin{figure}
\includegraphics[width=0.45\linewidth]{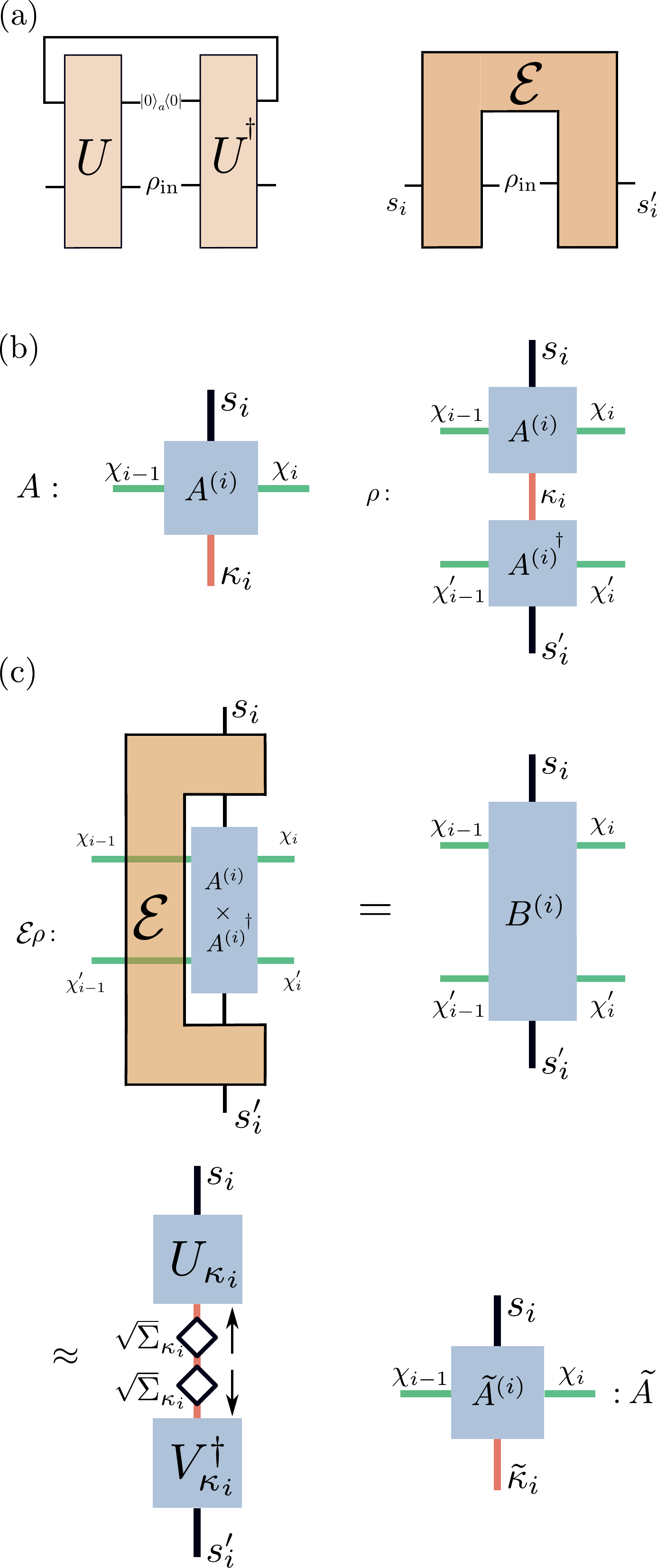}
\caption{(a) Action of the noise channel on an input density matrix $\rho_{\text{in}}$. 
(Left) Circuit representation of the noise channel. We couple the degrees of freedom
of the system with an ancilla qubit representing the environment. We evolve the coupled system 
with a unitary, $U$ finally tracing over the environmental degree of freedom resulting in a 
noise channel. (Right) An equivalent representation using the tensor network formalism.
$\mathcal{E}$ with two physical indices on the inside acting on the MPO representation of the 
input state, $\rho_{in}$, resulting in an updated state with two outer indices.
(b) To act with a channel on an LPDO, $A$, we first contract over the $\kappa_{i}$ index
resulting in an MPO, $\rho$. (c) The channel, $\mathcal{E}$ acts on $\rho$ wherein the inner
indices of $\mathcal{E}$ act on $s_{i}$ and $s'_{i}$ indices of $\rho$ resulting 
in an updated object $B^{(i)}$. We then perform a SVD in the $\kappa$-subspace and further 
truncate and renormalize the singular values, $\Sigma_{\kappa_{i}}$. Finally to return to the 
LPDO representation we integrate $\sqrt{\Sigma_{\kappa_{i}}}$ into the top and bottom 
tensors with the top tensor representing the updated LPDO, $\tilde{A}$. The action of single
body noise channel updates $\kappa_{i}$ to $\tilde{\kappa}_{i}$ while $\chi_{i-1}$
and $\chi_{i}$ remain the same.}
\label{fig:lpdo_channel}
\end{figure}

We discuss the application of the single-body noise channel on an LPDO. First, we briefly 
discuss the construction of the single-body noise channel followed by its application on
an LPDO. A noise channel is a completely positive trace preserving (CPTP) map obtained
by evolving the system and the environment followed by tracing out the environmental degrees
of freedom, see Fig.~\ref{fig:lpdo_channel}(a) for circuit representation and its 
tensor network equivalent. The noise channel action at a site $i$ of an LPDO involves 
contracting over the mixture index, $\kappa_{i}$ followed by the application of the 
noise channel. We then perform an SVD on the updated composite object in the 
$\kappa$-subspace to return to the LPDO representation, see Fig.~\ref{fig:lpdo_channel}(b).

\subsection{Isometric gauge freedom in the $\kappa$-subspace}
\begin{figure}
\includegraphics[width=0.55\linewidth]{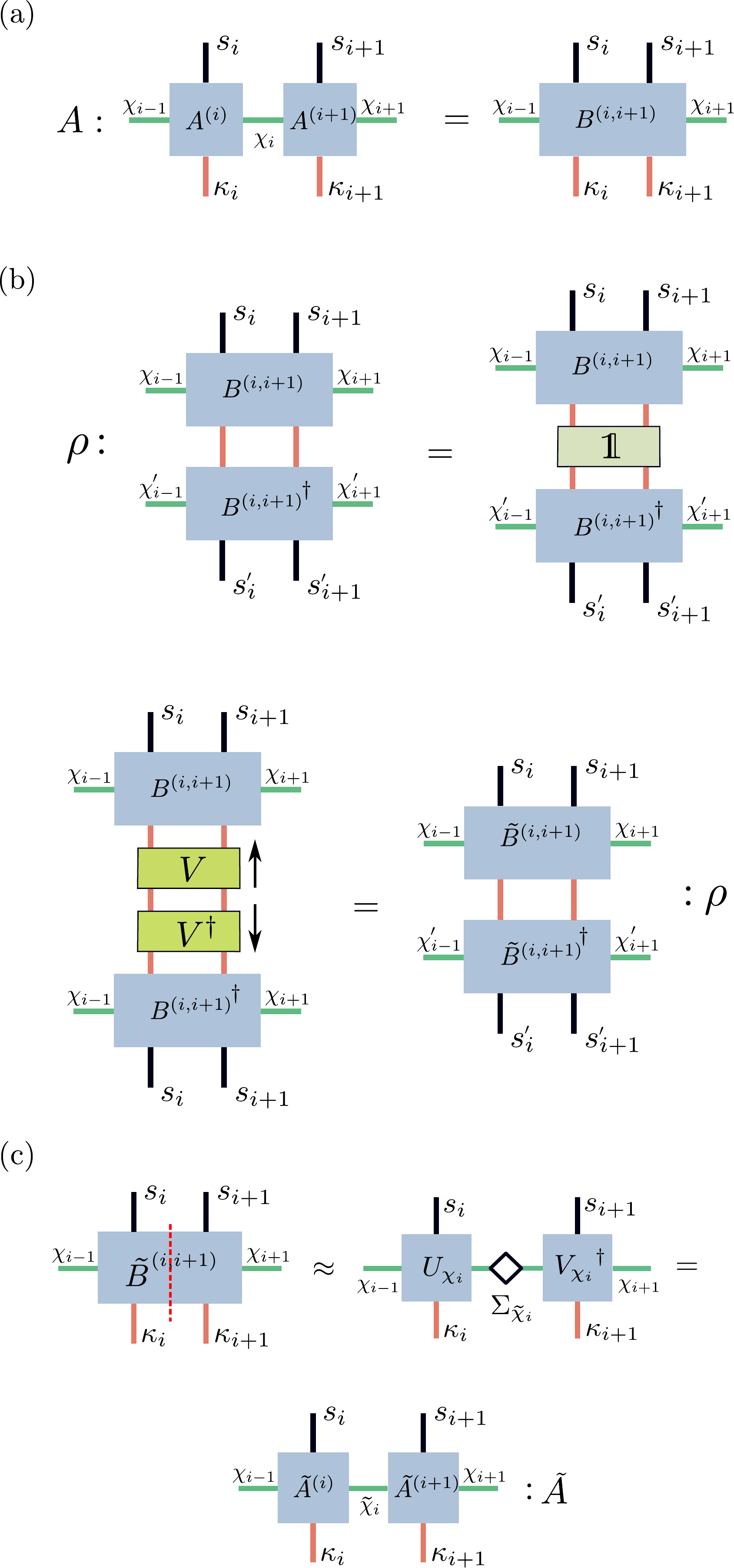}
\caption{Isometric guage freedom in the $\kappa$-subspace leading to two different
representations of the same density matrix. (a) Contracting over $\chi_{i}$ resulting
in $B^{(i, i+1)}$. (b) Constructing $\rho$ by contracting over the $\kappa_{i}$ and
$\kappa_{i+1}$ indices. Isometry, $V$ satisfying $VV^{\dag}=\mathds{1}$ can now be inserted 
into the $\kappa$-subspace. $V$ ($V^{\dag}$) are absorbed into $B^{(i,i+1)}$ 
(${B^{(i,i+1)}}^{\dag}$) resulting in $\tilde{B}^{(i,i+1)}$ (${{\tilde{B}}^{(i,i+1)\dag}}$),
an equivalent density matrix, $\rho$ with a different LPDO decomposition, 
$\tilde{B}\tilde{B}^{\dag}$. (d) To retrieve the updated LPDO due to the action of the isometry
$V$, we perform an SVD on the $\tilde{B}^{(i, i+1)}$ truncate and renormalize the singular 
values resulting in $\tilde{A}$ with updated tensors $\tilde{A}^{(i)}$, $\tilde{A}^{(i+1)}$
and $\tilde{\chi}$.}
\label{fig:lpdo_iso_gauge}
\end{figure}

In the context of MPSs, it is well established that there exists multiple representations for a
given quantum state. This is due to the fact that given a representation, one could generate
multiple equivalent representations by inserting isometries into the virtual $\chi$-indices.
In a similar spirit, for a given density matrix there exists multiple LPDO representations
that could be mapped between each other by the action of isometries in the $\kappa$-subspace.
In Fig.~\ref{fig:lpdo_iso_gauge} we illustrate the invariance of the density matrix
under the action of the isometries in the $\kappa$-subspace.

\section{Validating the different protocols\label{sec:at_validation}}

\begin{figure*}
\begin{center}
\begin{tabular}{cp{0.01mm}c}
\subfig{(a)}{\includegraphics[width=0.4\linewidth]{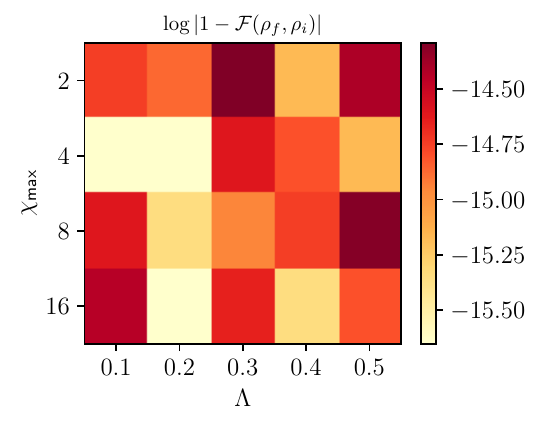}}
&&
\subfig{(b)}{\includegraphics[width=0.4\linewidth]{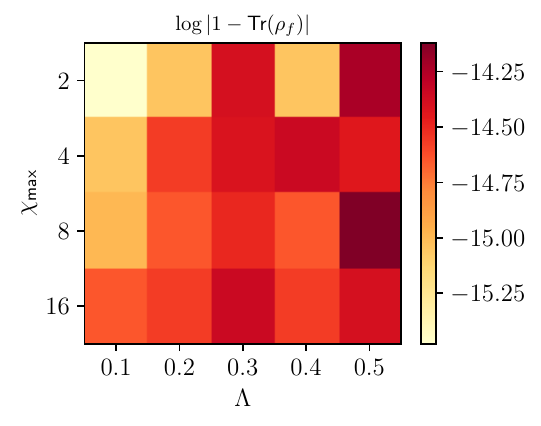}}\\
\subfig{(c)}{\includegraphics[width=0.4\linewidth]{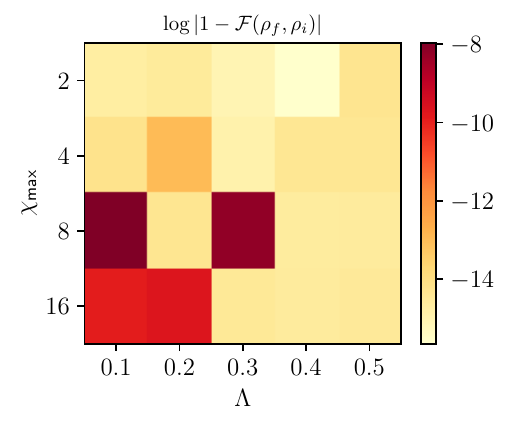}}
&&
\subfig{(d)}{\includegraphics[width=0.4\linewidth]{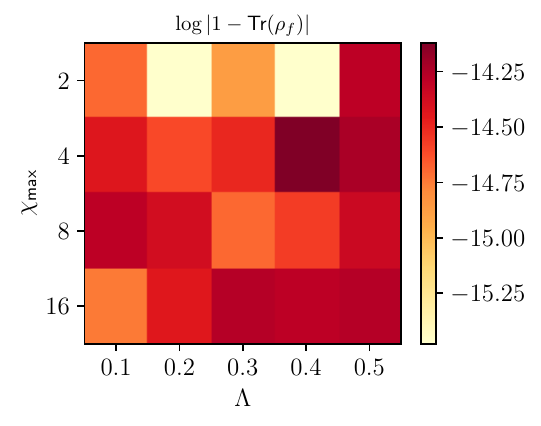}}\\
\subfig{(e)}{\includegraphics[width=0.4\linewidth]{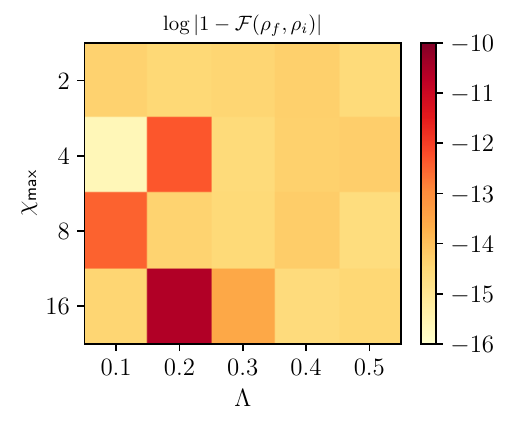}}
&&
\subfig{(f)}{\includegraphics[width=0.4\linewidth]{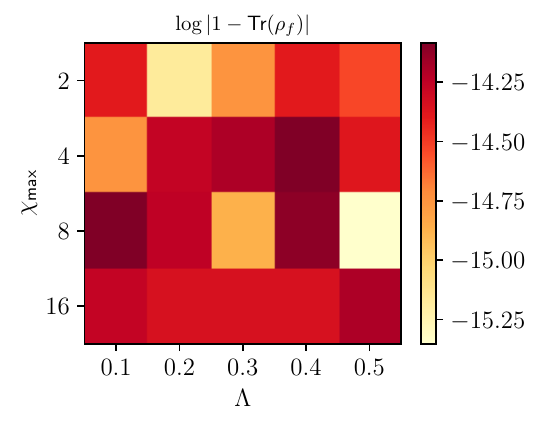}}\\
\end{tabular}
\end{center}
\caption{(a, c, e) Infidelity of the initial sub-optimal LPMM$\rho$ with the final optimized LPMM$\rho$ (the corresponding density matrices given by 
$\rho_{f}$ and $\rho_{i}$ respectively). (b, d, f) Deviation in norm of the final optimized LPMM$\rho$ from unity. Both, 
as a function of $\chi_\text{max}$, maximum bond dimension in the initial LPMM$\rho$ and $\chi$-cutoff, $\Lambda$ for a fixed 
system size of $N=100$. We employ in (a, b) fidelity-preserving truncation, (c-f) Reimannian manifold based optimization with objective functions
given by (c, d) $S_{\text{sr}}$, (e, f) $S_{\text{vn}}$. We observe that the infidelity measure in (c, e) obtained using the optimization routines
deviates significantly from unity, in the worst case scenario on the order of $10^{-5}$. This anomaly is predominantly observed for initial 
LPMM$\rho$ characterized by high $\chi_{\text{max}}$ and low cutoff values, $\Lambda$.}
\label{fig:at_chi_fid_norm}
\end{figure*}

In this appendix, we validate the different $\chi$-pruning protocols as introduced in the main text. To this 
extent, in Fig.~\ref{fig:at_chi_fid_norm}, we compute the fidelity, as in Eq.~\ref{eq:UJ}, 
\begin{equation}
 \label{eq:UJ}
     \mathcal{F}_P(\rho_{f}, \rho_{i}) = \frac{\text{Tr}[\rho_{f} \rho_{i}]}{\max(P(\rho_{f}),P(\rho_{i}))},
\end{equation}
where $P(\rho)$ represents the purity of $\rho$ and $\rho_{f} (\rho_{i})$ are the density matrices corresponding to the 
LPMM$\rho$'s post (pre) optimization. In addition, we also compute the trace of the optimized density matrix, 
$\text{Tr}[\rho_{f}]$ and observe that the norm remains conserved for all the $\chi$-pruning protocols. However, 
while the fidelity remains unity up to machine precision in the context of fidelity-preserving truncation protocol, in 
some scenarios we see a significant deviation (on the order of $10^{-5}$ in the worst case scenario away 
from unity) when the optimization protocol is employed.

\section{Exponential scaling in the fidelity-preserving truncation protocol\label{sec:scaling_at}}
\begin{figure*}
\begin{center}
\begin{tabular}{cp{0.01mm}cp{0.01mm}c}
\subfig{(a)}{\includegraphics[width=0.29\linewidth]{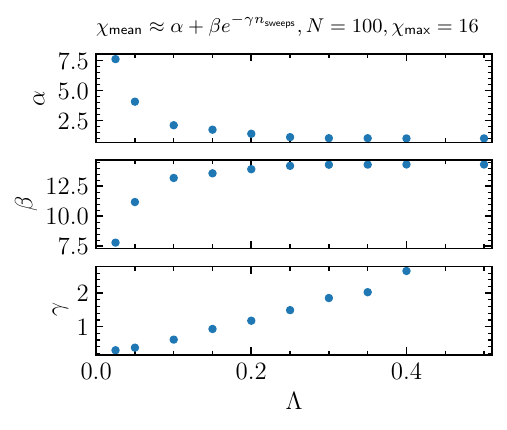}}
&&
\subfig{(a)}{\includegraphics[width=0.29\linewidth]{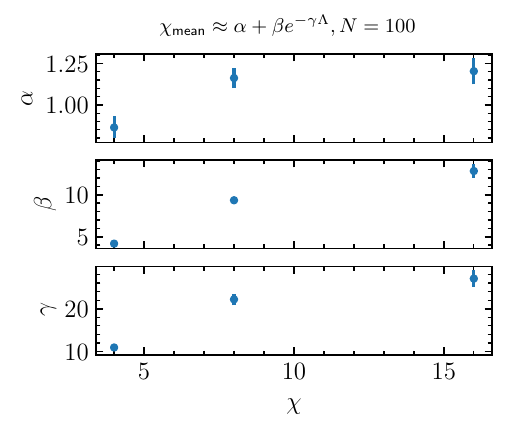}}
&&
\subfig{(c)}{\includegraphics[width=0.29\linewidth]{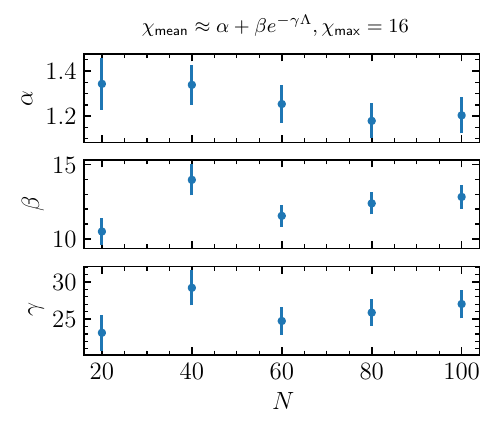}}
\end{tabular}
\end{center}
\caption{Exponential fit parameters: ($\alpha, \beta, \gamma$) for (a) $\chi_{\text{mean}}$ as an exponential function of $n_\text{sweeps}$ for an LPMM$\rho$ of size $N=100$ characterized by $\chi_{\text{max}}=16$ for different cutoffs $\Lambda$. (b) $\chi_{\text{mean}}$ as an
exponential function of cutoff $\Lambda$ for an LPMM$\rho$ of size $N=100$ for different $\chi$. (c) $\chi_{\text{mean}}$ as an exponential 
function of cutoff, $\Lambda$ for fixed $\chi_{\text{max}} = 16$ for different system sizes $N$. The error bars of the fit parameters 
are the standard deviation obtained from the diagonal of the covariance matrix associated with the fit.}
\label{fig:at_fit_coeffs}
\end{figure*}

In Fig.~\ref{fig:at_chi_avg_vs_cutoff}, we observe that by employing the fidelity-preserving truncation protocol 
the averaged $\chi_{\text{mean}}$ scales exponentially as a function of sweeps as well as the cutoff, $\Lambda$
for a given $N$ and $\chi_{\text{max}}$. To quantify the above relationships, we fit the data to an exponential
function $f(x) = \alpha + \beta e^{-\gamma x}$ and present the fit parameters as in Fig.~\ref{fig:at_fit_coeffs}.

\section{Symmetry transformations on the LPDO~\label{sec:app_d}}

In this section, we briefly review the key concepts involving the symmetry operations on LPDO. We note that most of the presentation is based on Ref.~\onlinecite{Guo2025} and we direct the interested reader to the above reference for a more detailed discussion. We begin by reviewing the notion of strong and weak symmetries for a given density matrix, $\rho$. 

\begin{definition}
    Unitary, $U$ is said to be a strong symmetry of the density matrix $\rho$ if $\rho$ remains invariant under the action of $U$ from the left, i.e., $U\rho = e^{i\phi}\rho$ for some phase $\phi$.
\end{definition}

In the context of the LDPO, a strong symmetry is represented by
\begin{equation}
\includegraphics[width=0.6\linewidth]{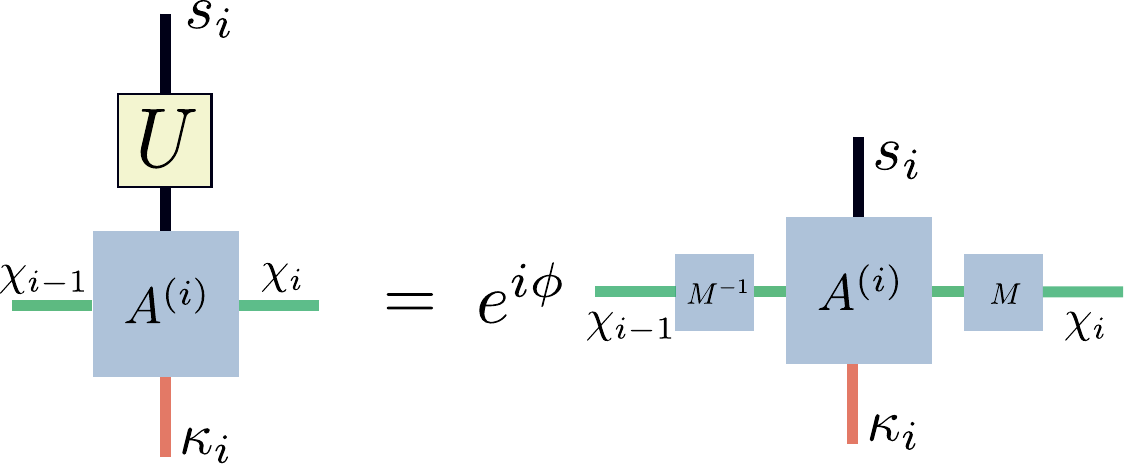} 
\end{equation}
That is, the representation remains invariant upto some gauge freedom in the $\chi$-space under the action of a unitary, $U$ with the isometric gauge freedom in the mixture space set to identity. Intuitively, in the case that the unitary, $U$ (acting only from the left) induces an non-identity isometry, the resulting $\rho$ obtained by contracting over the mixture index, would no longer be invariant under the unitary. 

\begin{definition}
    Unitary, $U$ is said to be a weak symmetry of the density matrix $\rho$ if $\rho$ remains invariant under the conjugation of $U$ i.e., $U\rho U^{\dag} = \rho$.
\end{definition}


This in the LPDO representation translates to
\begin{equation}
\includegraphics[width=0.6\linewidth]{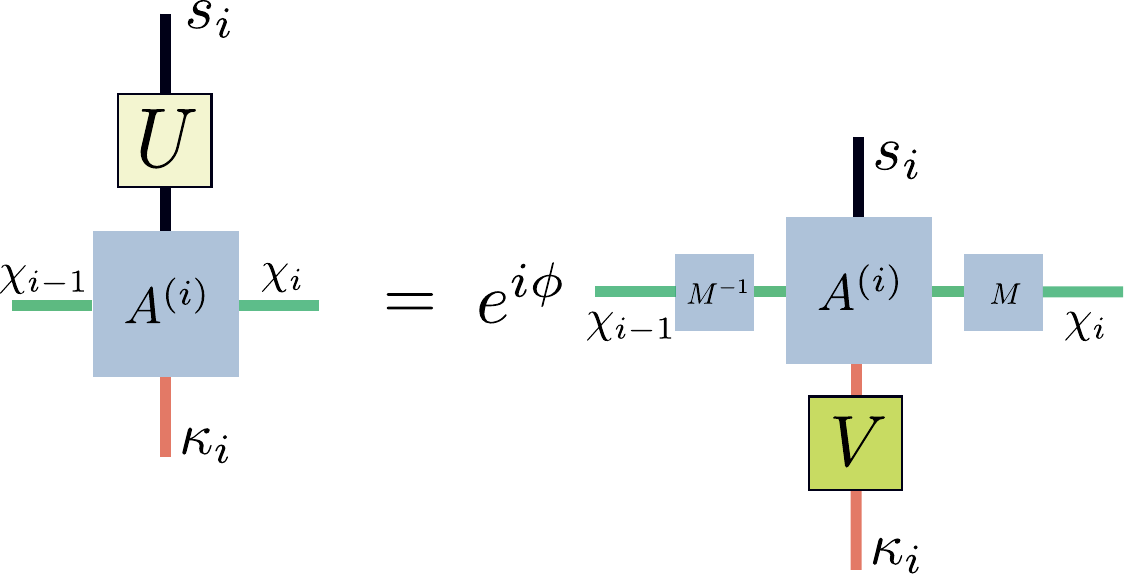} 
\end{equation}
The representation remains invariant upto some gauge freedom in the $\chi$-space and some isometry $V$ in the mixture space under the action of unitary $U$.

\bibliography{bib}

@misc{Jamadagni2024,
      title={Gauge-Fixing Quantum Density Operators At Scale}, 
      author={Amit Jamadagni and Eugene Dumitrescu},
      year={2024},
      eprint={2411.03548},
      archivePrefix={arXiv},
      primaryClass={quant-ph},
      url={https://arxiv.org/abs/2411.03548}, 
}

@misc{Le2025,
      title={Riemannian quantum circuit optimization based on matrix product operators}, 
      author={Isabel Nha Minh Le and Shuo Sun and Christian B. Mendl},
      year={2025},
      eprint={2501.08872},
      archivePrefix={arXiv},
      primaryClass={quant-ph},
      url={https://arxiv.org/abs/2501.08872}, 
}

@misc{Rogerson2024,
      title={Quantum Circuit Optimization using Differentiable Programming of Tensor Network States}, 
      author={David Rogerson and Ananda Roy},
      year={2024},
      eprint={2408.12583},
      archivePrefix={arXiv},
      primaryClass={quant-ph},
      url={https://arxiv.org/abs/2408.12583}, 
}

@misc{Ramirez2024,
      title={A Riemannian Approach to the Lindbladian Dynamics of a Locally Purified Tensor Network}, 
      author={Emiliano Godinez-Ramirez and Richard Milbradt and Christian B. Mendl},
      year={2024},
      eprint={2409.08127},
      archivePrefix={arXiv},
      primaryClass={quant-ph},
      url={https://arxiv.org/abs/2409.08127}, 
}

@article{Zygote2018,
  author    = {Michael Innes},
  title     = {Don't Unroll Adjoint: Differentiating SSA-Form Programs},
  journal   = {CoRR},
  volume    = {abs/1810.07951},
  year      = {2018},
  url       = {http://arxiv.org/abs/1810.07951},
  archivePrefix = {arXiv},
  eprint    = {1810.07951},
  bibsource = {dblp computer science bibliography, https://dblp.org}
}

@article{Evenbly2009,
   title={Algorithms for entanglement renormalization},
   volume={79},
   ISSN={1550-235X},
   url={http://dx.doi.org/10.1103/PhysRevB.79.144108},
   DOI={10.1103/physrevb.79.144108},
   number={14},
   journal={Physical Review B},
   publisher={American Physical Society (APS)},
   author={Evenbly, G. and Vidal, G.},
   year={2009},
   month=apr 
}

@article{Bergmann2022,
    Author    = {Ronny Bergmann},
    Doi       = {10.21105/joss.03866},
    Journal   = {Journal of Open Source Software},
    Number    = {70},
    Pages     = {3866},
    Publisher = {The Open Journal},
    Title     = {Manopt.jl: Optimization on Manifolds in {J}ulia},
    Volume    = {7},
    Year      = {2022},
}

@article{Axen2023,
    AUTHOR    = {Axen, Seth D. and Baran, Mateusz and Bergmann, Ronny and Rzecki, Krzysztof},
    ARTICLENO = {33},
    DOI       = {10.1145/3618296},
    JOURNAL   = {ACM Transactions on Mathematical Software},
    MONTH     = {dec},
    NUMBER    = {4},
    TITLE     = {Manifolds.Jl: An Extensible Julia Framework for Data Analysis on Manifolds},
    VOLUME    = {49},
    YEAR      = {2023}
}

@Article{Kevin2021,
	title={{Fast tensor disentangling algorithm}},
	author={Kevin Slagle},
	journal={SciPost Phys.},
	volume={11},
	pages={056},
	year={2021},
	publisher={SciPost},
	doi={10.21468/SciPostPhys.11.3.056},
	url={https://scipost.org/10.21468/SciPostPhys.11.3.056},
}

@article{Werner2016,
   title={Positive Tensor Network Approach for Simulating Open Quantum Many-Body Systems},
   volume={116},
   ISSN={1079-7114},
   url={http://dx.doi.org/10.1103/PhysRevLett.116.237201},
   DOI={10.1103/physrevlett.116.237201},
   number={23},
   journal={Physical Review Letters},
   publisher={American Physical Society (APS)},
   author={Werner, A. H. and Jaschke, D. and Silvi, P. and Kliesch, M. and Calarco, T. and Eisert, J. and Montangero, S.},
   year={2016},
   month=jun }

@article{Hauschild2018,
   title={Finding purifications with minimal entanglement},
   volume={98},
   ISSN={2469-9969},
   url={http://dx.doi.org/10.1103/PhysRevB.98.235163},
   DOI={10.1103/physrevb.98.235163},
   number={23},
   journal={Physical Review B},
   publisher={American Physical Society (APS)},
   author={Hauschild, Johannes and Leviatan, Eyal and Bardarson, Jens H. and Altman, Ehud and Zaletel, Michael P. and Pollmann, Frank},
   year={2018},
   month=dec }

@article{Hauru2021,
   title={Riemannian optimization of isometric tensor networks},
   volume={10},
   ISSN={2542-4653},
   url={http://dx.doi.org/10.21468/SciPostPhys.10.2.040},
   DOI={10.21468/scipostphys.10.2.040},
   number={2},
   journal={SciPost Physics},
   publisher={Stichting SciPost},
   author={Hauru, Markus and Van Damme, Maarten and Haegeman, Jutho},
   year={2021},
   month=feb 
}

@article{Guo2025,
   title={Locally Purified Density Operators for Symmetry-Protected Topological Phases in Mixed States},
   volume={15},
   ISSN={2160-3308},
   url={http://dx.doi.org/10.1103/PhysRevX.15.021060},
   DOI={10.1103/physrevx.15.021060},
   number={2},
   journal={Physical Review X},
   publisher={American Physical Society (APS)},
   author={Guo, Yuchen and Zhang, Jian-Hao and Zhang, Hao-Ran and Yang, Shuo and Bi, Zhen},
   year={2025},
   month=may 
}

@misc{Wanisch2025,
      title={Entanglement transitions in a boundary-driven open quantum many-body system}, 
      author={Darvin Wanisch and Nora Reinić and Daniel Jaschke and Simone Montangero and Pietro Silvi},
      year={2025},
      eprint={2502.18372},
      archivePrefix={arXiv},
      primaryClass={quant-ph},
      url={https://arxiv.org/abs/2502.18372}, 
}

@article{Harada2025,
  doi = {10.22331/q-2025-08-07-1823},
  url = {https://doi.org/10.22331/q-2025-08-07-1823},
  title = {Density matrix representation of hybrid tensor networks for noisy quantum devices},
  author = {Harada, Hiroyuki and Suzuki, Yasunari and Yang, Bo and Tokunaga, Yuuki and Endo, Suguru},
  journal = {{Quantum}},
  issn = {2521-327X},
  publisher = {{Verein zur F{\"{o}}rderung des Open Access Publizierens in den Quantenwissenschaften}},
  volume = {9},
  pages = {1823},
  month = aug,
  year = {2025}
}

@article{Verstraete2004_mpdo,
   title={Matrix Product Density Operators: Simulation of Finite-Temperature and Dissipative Systems},
   volume={93},
   ISSN={1079-7114},
   url={http://dx.doi.org/10.1103/PhysRevLett.93.207204},
   DOI={10.1103/physrevlett.93.207204},
   number={20},
   journal={Physical Review Letters},
   publisher={American Physical Society (APS)},
   author={Verstraete, F. and García-Ripoll, J. J. and Cirac, J. I.},
   year={2004},
   month=nov }

@misc{Mueller2024,
      title={Enabling large-depth simulation of noisy quantum circuits with positive tensor networks}, 
      author={Ambroise Müller and Thomas Ayral and Corentin Bertrand},
      year={2024},
      eprint={2403.00152},
      archivePrefix={arXiv},
      primaryClass={quant-ph},
      url={https://arxiv.org/abs/2403.00152}, 
}

@misc{Cichy2025,
    title={Classical simulation of noisy quantum circuits via locally entanglement-optimal unravelings},
    author={Simon Cichy and Paul K. Faehrmann and Lennart Bittel and Jens Eisert and Hakop Pashayan},
    year={2025},
    eprint={2508.05745},
    archivePrefix={arXiv},
    primaryClass={quant-ph}
}

@misc{Perez2007,
      title={Matrix Product State Representations}, 
      author={D. Perez-Garcia and F. Verstraete and M. M. Wolf and J. I. Cirac},
      year={2007},
      eprint={quant-ph/0608197},
      archivePrefix={arXiv},
      primaryClass={quant-ph},
      url={https://arxiv.org/abs/quant-ph/0608197}, 
}

@misc{Verstraete2004,
      title={Renormalization algorithms for Quantum-Many Body Systems in two and higher dimensions}, 
      author={F. Verstraete and J. I. Cirac},
      year={2004},
      eprint={cond-mat/0407066},
      archivePrefix={arXiv},
      primaryClass={cond-mat.str-el},
      url={https://arxiv.org/abs/cond-mat/0407066}, 
}

@article{Vidal2007,
   title={Entanglement Renormalization},
   volume={99},
   ISSN={1079-7114},
   url={http://dx.doi.org/10.1103/PhysRevLett.99.220405},
   DOI={10.1103/physrevlett.99.220405},
   number={22},
   journal={Physical Review Letters},
   publisher={American Physical Society (APS)},
   author={Vidal, G.},
   year={2007},
   month=nov }

@misc{Thompson2025,
      title={Non-zero noise extrapolation: accurately simulating noisy quantum circuits with tensor networks}, 
      author={Anthony P. Thompson and Arie Soeteman and Chris Cade and Ido Niesen},
      year={2025},
      eprint={2501.13237},
      archivePrefix={arXiv},
      primaryClass={quant-ph},
      url={https://arxiv.org/abs/2501.13237}, 
}

@article{Li2017,
   title={Efficient Variational Quantum Simulator Incorporating Active Error Minimization},
   volume={7},
   ISSN={2160-3308},
   url={http://dx.doi.org/10.1103/PhysRevX.7.021050},
   DOI={10.1103/physrevx.7.021050},
   number={2},
   journal={Physical Review X},
   publisher={American Physical Society (APS)},
   author={Li, Ying and Benjamin, Simon C.},
   year={2017},
   month=jun }

@article{Temme2017,
   title={Error Mitigation for Short-Depth Quantum Circuits},
   volume={119},
   ISSN={1079-7114},
   url={http://dx.doi.org/10.1103/PhysRevLett.119.180509},
   DOI={10.1103/physrevlett.119.180509},
   number={18},
   journal={Physical Review Letters},
   publisher={American Physical Society (APS)},
   author={Temme, Kristan and Bravyi, Sergey and Gambetta, Jay M.},
   year={2017},
   month=nov }

@article{Arute2019,
   title={Quantum supremacy using a programmable superconducting processor},
   volume={574},
   ISSN={1476-4687},
   url={http://dx.doi.org/10.1038/s41586-019-1666-5},
   DOI={10.1038/s41586-019-1666-5},
   number={7779},
   journal={Nature},
   publisher={Springer Science and Business Media LLC},
   author={Arute, Frank and Arya, Kunal and Babbush, Ryan and Bacon, Dave and others},
   year={2019},
   month=oct, pages={505–510} }

@article{Kandala2018,
  title={Error mitigation extends the computational reach of a noisy quantum processor},
  author={Abhinav Kandala and Kristan Temme and Antonio D. C{\'o}rcoles and Antonio Mezzacapo and Jerry M. Chow and Jay M. Gambetta},
  journal={Nature},
  year={2018},
  volume={567},
  pages={491 - 495},
  url={https://doi.org/10.1038/s41586-019-1040-7},
  DOI={10.1038/s41586-019-1040-7},
}

@article{Cai2021,
   title={Multi-exponential error extrapolation and combining error mitigation techniques for NISQ applications},
   volume={7},
   ISSN={2056-6387},
   url={http://dx.doi.org/10.1038/s41534-021-00404-3},
   DOI={10.1038/s41534-021-00404-3},
   number={1},
   journal={npj Quantum Information},
   publisher={Springer Science and Business Media LLC},
   author={Cai, Zhenyu},
   year={2021},
   month=may 
}

@article{Frey2022,
author = {Philipp Frey  and Stephan Rachel },
title = {Realization of a discrete time crystal on 57 qubits of a quantum computer},
journal = {Science Advances},
volume = {8},
number = {9},
pages = {eabm7652},
year = {2022},
doi = {10.1126/sciadv.abm7652},
URL = {https://www.science.org/doi/abs/10.1126/sciadv.abm7652},
eprint = {https://www.science.org/doi/pdf/10.1126/sciadv.abm7652}
}

@article{Moharramipour2024,
   title={Symmetry-Enforced Entanglement in Maximally Mixed States},
   volume={5},
   ISSN={2691-3399},
   url={http://dx.doi.org/10.1103/PRXQuantum.5.040336},
   DOI={10.1103/prxquantum.5.040336},
   number={4},
   journal={PRX Quantum},
   publisher={American Physical Society (APS)},
   author={Moharramipour, Amin and Lessa, Leonardo A. and Wang, Chong and Hsieh, Timothy H. and Sahu, Subhayan},
   year={2024},
   month=dec }

@misc{Chen2024,
      title={Unconventional topological mixed-state transition and critical phase induced by self-dual coherent errors}, 
      author={Yu-Hsueh Chen and Tarun Grover},
      year={2024},
      eprint={2403.06553},
      archivePrefix={arXiv},
      primaryClass={quant-ph},
      url={https://arxiv.org/abs/2403.06553}, 
}

@article{Ellison2025,
   title={Toward a Classification of Mixed-State Topological Orders in Two Dimensions},
   volume={6},
   ISSN={2691-3399},
   url={http://dx.doi.org/10.1103/PRXQuantum.6.010315},
   DOI={10.1103/prxquantum.6.010315},
   number={1},
   journal={PRX Quantum},
   publisher={American Physical Society (APS)},
   author={Ellison, Tyler D. and Cheng, Meng},
   year={2025},
   month=jan }

@article{Nguyen2018,
   title={Entanglement of purification: from spin chains to holography},
   volume={2018},
   ISSN={1029-8479},
   url={http://dx.doi.org/10.1007/JHEP01(2018)098},
   DOI={10.1007/jhep01(2018)098},
   number={1},
   journal={Journal of High Energy Physics},
   publisher={Springer Science and Business Media LLC},
   author={Nguyen, Phuc and Devakul, Trithep and Halbasch, Matthew G. and Zaletel, Michael P. and Swingle, Brian},
   year={2018},
   month=Jan }

@article{Zhou2020,
   title={What Limits the Simulation of Quantum Computers?},
   volume={10},
   ISSN={2160-3308},
   url={http://dx.doi.org/10.1103/PhysRevX.10.041038},
   DOI={10.1103/physrevx.10.041038},
   number={4},
   journal={Physical Review X},
   publisher={American Physical Society (APS)},
   author={Zhou, Yiqing and Stoudenmire, E. Miles and Waintal, Xavier},
   year={2020},
   month=Nov }

@article{Cuevas2013,
   title={Purifications of multipartite states: limitations and constructive methods},
   volume={15},
   ISSN={1367-2630},
   url={http://dx.doi.org/10.1088/1367-2630/15/12/123021},
   DOI={10.1088/1367-2630/15/12/123021},
   number={12},
   journal={New Journal of Physics},
   publisher={IOP Publishing},
   author={Cuevas, Gemma De las and Schuch, Norbert and Pérez-García, David and Ignacio Cirac, J},
   year={2013},
   month=Dec, pages={123021} }

@article{Jaschke2019,
  title={One-dimensional many-body entangled open quantum systems with tensor network methods},
  author={Jaschke, Daniel and Montangero, Simone and Carr, Lincoln D},
  journal={Quantum science and technology},
  volume={4},
  number={1},
  pages={013001},
  year={2019},
  publisher={IOP Publishing}
}

@article{Surace2019,
   title={Simulating the out-of-equilibrium dynamics of local observables by trading entanglement for mixture},
   volume={99},
   ISSN={2469-9969},
   url={http://dx.doi.org/10.1103/PhysRevB.99.235115},
   DOI={10.1103/physrevb.99.235115},
   number={23},
   journal={Physical Review B},
   publisher={American Physical Society (APS)},
   author={Surace, J. and Piani, M. and Tagliacozzo, L.},
   year={2019},
   month=June }

@article{Cuevas2020,
   title={Mixed states in one spatial dimension: Decompositions and correspondence with nonnegative matrices},
   volume={61},
   ISSN={1089-7658},
   url={http://dx.doi.org/10.1063/1.5127668},
   DOI={10.1063/1.5127668},
   number={4},
   journal={Journal of Mathematical Physics},
   publisher={AIP Publishing},
   author={de las Cuevas, Gemma and Netzer, Tim},
   year={2020},
   month=Apr }

@article{Guo2024,
   title={Locally purified density operators for noisy quantum circuits},
   volume={41},
   ISSN={1741-3540},
   url={http://dx.doi.org/10.1088/0256-307X/41/12/120302},
   DOI={10.1088/0256-307x/41/12/120302},
   number={12},
   journal={Chinese Physics Letters},
   publisher={IOP Publishing},
   author={Guo, Yuchen and Yang, Shuo},
   year={2024},
   month=Dec, pages={120302} }

@article{Guo2022,
   title={Density matrix renormalization group algorithm for mixed quantum states},
   volume={105},
   ISSN={2469-9969},
   url={http://dx.doi.org/10.1103/PhysRevB.105.195152},
   DOI={10.1103/physrevb.105.195152},
   number={19},
   journal={Physical Review B},
   publisher={American Physical Society (APS)},
   author={Guo, Chu},
   year={2022},
   month=May }

\end{document}